\documentclass[manuscript]{acmart}
\usepackage[modulo]{lineno}
\AtBeginDocument{%
  }
\setcopyright{acmlicensed}
\copyrightyear{2026}
\acmYear{2026}
\graphicspath{{images/}}

\begin{document} 
\nolinenumbers

\title{BioMetaphor: AI-Generated Biodata Representations for Virtual Co-Present Events}


\author{Lin Lin}
\affiliation{%
  \institution{School of Design, South University of Science and Technology of China}
  \city{Shenzhen}
  \country{China}}
\email{12431501@mail.sustech.edu.cn}

\author{Ming Wu}
\affiliation{%
  \institution{School of Design, South University of Science and Technology of China}
  \city{Shenzhen}
  \country{China}}
\email{12531632@mail.sustech.edu.cn}

\author{Anyu Ren}
\affiliation{%
  \institution{School of Design, South University of Science and Technology of China}
  \city{Shenzhen}
  \country{China}}
\email{raynkkk@163.com}

\author{Zhanwei Wu}
\affiliation{%
  \institution{School of Design, Shanghai Jiao Tong University}
  \city{Shanghai}
  %\state{Beijing Shi}
  \country{China}}
\email{zhanwei_wu@sjtu.edu.cn}

\author{Daojun GONG}
\affiliation{%
  \institution{Northwestern Polytechnical University}
  \city{Shenzhen}
  \country{China}}
\email{iamdjgong@mail.nwpu.edu.cn}

\author{Ruowei Xiao}
\affiliation{%
 \institution{School of Design, South University of Science and Technology of China}
 \city{Shenzhen}
 \country{China}}
 \email{xiaorw@sustech.edu.cn}

%\renewcommand{\shortauthors}{L Lin et al.}

\begin{abstract}
In virtual or hybrid co-present events, biodata is emerging as a new paradigm of social cues. While it is able to reveal individuals' inner states, the technology-mediated representation of biodata in social contexts remains underexplored. This study aims to uncover human cognitive preferences and patterns for biodata expression and leverage this knowledge to guide generative AI (GenAI) in creating biodata representations for co-present experiences, aligning with the broader concept of Human-in-the-loop. We conducted a user elicitation workshop with 30 HCI experts and investigated the results using qualitative analysis. Based on our findings, we further propose a GenAI-driven framework: BioMetaphor. Our framework demonstration shows that current GenAI can learn and express visual biodata cues in an event-adpated, human-like manner. This human-centered approach engages users in research, revealing the underlying cognition constructions for biodata expression while demonstrating how such knowledge can inform the design and development of future empathic technologies.
\end{abstract}



  %\includegraphics[width=\textwidth]{sampleteaser}
  %\caption{Seattle Mariners at Spring Training, 2010.}
  %\Description{Enjoying the baseball game from the third-base
  %seats. Ichiro Suzuki preparing to bat.}
  %\label{fig:teaser}

\begin{teaserfigure}
  \centering
  \includegraphics[width=\textwidth]{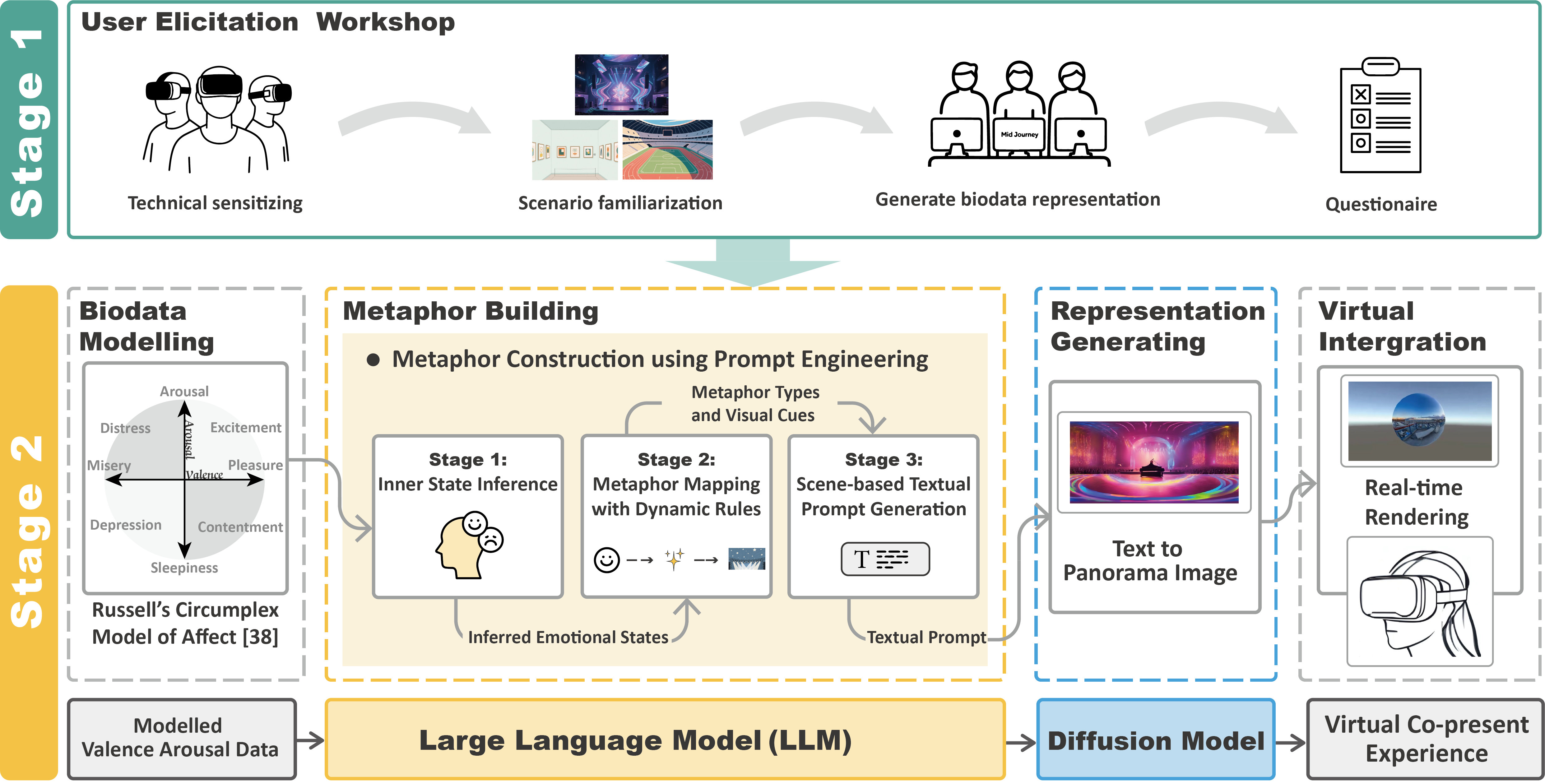}
  \caption{Overview of the research structure and GenAI framework for biosensory information representations} 
  \Description{The image is a two-stage diagram illustrating a process for biodata representation and virtual experience generation.
  Stage 1: user elicitation Workshop. This stage consists of four sequential steps: 1) Technical Sensitizing: Depicts three people wearing virtual reality (VR) headsets. 2) Scenario Familiarization: Displays images of three virtual environments: a concert, a gallery, and a sports field. 3) Generate Biodata Representation with GenAI: Illustrates three people sitting at a desk with a laptop, displaying a Mid-journey interface. 4) Questionnaire: Features a clipboard with options. 
  Stage 2: Biodata Representation and Virtual Integration.This stage is divided into four main modules: 
  1) Biodata Modelling Module: Includes a circular diagram labeled "Russell’s Circumplex Model of Affect." This is a two-dimensional graph with the x-axis labeled "Valence" (ranging from "Misery" on the left to "Pleasure" on the right) and the y-axis labeled "Arousal" (ranging from "Sleepiness" at the bottom to "Arousal" at the top). Specific emotional states are marked, such as "Distress" (high arousal, low valence), "Contentment" (low arousal, high valence), and "Sleepiness" (low arousal, low valence). 
  2) Metaphor Building Module: Titled "Metaphor Construction using Prompt Engineering", this section is divided into three stages: Stage 1: Inner State Inference: Shows a silhouette of a head with a smiling face and a sad face. An arrow labeled "Inferred Emotional States" connects stage 1 to stage 2. Stage 2: Metaphor Mapping with Dynamic Rules: Depicts a combination of the smiling face with the dynamic symbols. An arrow labeled "Metaphor Types and Visual Cues" connects stage 2 to stage 3.
  Stage 3：Scene-based Textual Prompt Generation: Shows a text box. An arrow labeled "textual prompt" points from Stage 3 to next module.
  3) Representation Generation Module: Displays a panoramic image of a virtual concert scene with vibrant pink and purple lights, abstract shapes, and a reflective floor, labeled "Text to Panorama Image." This section is connected to the "Metaphor Building" section via an arrow.
  4) Virtual Integration Module: Includes an image of a virtual environment with a spherical object showing the rendering process in 3D engine, labeled "Real-time Rendering." Below the text, a people wearing a VR headset is shown.
  At the bottom of Stage 2, a flowchart illustrates the progression from "Modelled Valence Arousal Data" (corresponding to the Biodata Modelling Module), to a "Large Language Model (LLM)" (corresponding to the Metaphor Building Module), then to a "Diffusion Model" (corresponding to the Representation Generation Module), and finally to the "Virtual Co-present Experience" (corresponding to the Virtual Integration Module).
  }
  \label{fig:teaser.png}
\end{teaserfigure}


\maketitle

\section{Introduction}
In fully or partially virtual events, online (and offline) users can be co-present and share immersive experiences, such as the COLDPLAY VR concert \cite{andrews2020concerts}, the Online IEEE VR Conference 2021 \cite{moreira2022toward}, etc. This kind of virtual or hybrid co-present events, however, still exposes a significant gap when compared to its physical counterpart. Currently, avatar-mediated communication prevails in such venues, emphasizing one-to-one replication of explicit social cues, e.g. verbal language, facial expression and gestures etc. Prior research suggests that the paradigm based on virtual avatars has inherent problems such as the uncanny valley effect \cite{shin_uncanny_2019}, stereotype issues \cite{curran_understanding_2019} and social abuse \cite{wang_systematic_2024}. Specifically, in many large-scale co-present scenes, crowd and collective status (e.g., group-level arousal and valence \cite{mou2015group}, group-level behavioral synchrony \cite{ hove2009s, valdesolo2011synchrony}) are considered as more "information of interest", where avatars may not be an innately coherent and cost-effective representation. Alternative to the avatar-based paradigm, HCI research community now witnesses a growing interests in exploring implicit social cues instead, such as biosensory information, that reveal and reflect individuals' internal psychophysiological states \cite{wang_systematic_2024}. 

Heartbeats \cite{hirsch2023my}, respiration rate \cite{salminen_evoking_2019}, electroencephalogram (EEG) \cite{souza2021attention} and other kinds of biodata have already been widely applied to technology-mediated social interaction and information sharing in existing literature. With proper design, this approach is believed to reduce biases associated with race, gender and class, thereby potentially preserving social norms and promoting equality \cite{gong_comparing_2025}. While previous studies intensively concentrated on  biodata collection, processing and classification \cite{halbig_systematic_2021}, less research is found on particular ways of presenting and expressing biodata. Plain data visualization \cite{hirsch2023my}, sonification \cite{winters_can_2021} and haptification \cite{dey_effects_2018} were the major approaches found in current research concerning biosensory information representation, which, in essence, are all linear mappings of raw biodata. Some researchers have also started exploring the use of different visual imagery, e.g. heart icon, creature and ambient light, as meaningful social cues in VR context \cite{lee_understanding_2022}. Yet, fewer studies been done to systematically investigate users’ cognitive preferences and its underlying mechanisms regarding biodata representation. In particular, different co-present events often have varying and situational user needs for expressing biodata \cite{gong_comparing_2025,lux2018live}, which also raises specific requirements on the methods and tools for constructing such representations. 

The rapid advancement of generative artificial intelligence (GenAI) offers a promising path in this field. Large language models (LLMs), diffusion models and other alike GenAIs are trained based on massive human-generated contents, e.g. text corpus, tagged image database etc. Arguably, they reflect crowd or collective human knowledge on average, at least in a statistical sense  \cite{Bender_Gebru_McMillan-Major_Shmitchell_2021, lake2017building}. However, current GenAIs have not yet been sufficiently trained to understand or explain human biodata and the rich information it reveals. While it is among the inevitable trends in the evolution toward next-generation multimodal GenAI, it entails tremendous human engagement and intervention to "teach" GenAIs how to analyze, reason and ultimately express biosensory information in a human-preferred manner. In this study, we posed the following research questions (RQs):
\begin{itemize}
    \item \textbf{RQ1.} What do people prefer for expressing biodata in co-present events?
    \item \textbf{RQ2.} How can GenAI be used to generate biodata representations that align with human preferences?
\end{itemize}

We incorporate insights from the Human-in-the-Loop (HITL) \cite{Dautenhahn_1998} approach to elicit users' preferences for biodata representation and use it to guide GenAI models. Explicitly, for RQ1 we conducted a user elicitation workshop with 30 HCI experts and invited them to create their own biodata expressions by using GenAI tools within three representative co-present scenarios, respectively VR gallery, sports event and concert. We investigated several aspects of user preferences by qualitatively analyzing the workshop results: representation modality, level of interpretation, and level of understanding. One of our major observations revealed that participants’ outcomes generally aligned with a metaphorical frame of visual biodata representations. These findings hence led to our solution for RQ2: We employed prompt engineering to reflect the knowledge and insights extracted in the previous stage and proposed a GenAI-driven framework, \textit{BioMetaphor}. This framework leverages a multi-GenAI approach to infer users' internal states based on modeled biodata, construct corresponding metaphorical expressions, and finally render them into visual social cues in co-present scenarios, as shown in Fig.\ref{fig:teaser.png}.   

The contributions of this work are: 1) Through a qualitative analysis of the user elicitation workshop results, we revealed human preferences for understanding and expressing biodata in co-present events. 2) We proposed \textbf{BioMetaphor}, a GenAI-driven framework to express biodata in line with human cognitive preferences. 3) We demonstrated the current GenAI's ability to learn and, to some extent, imitate advanced human cognitive patterns. We expect this work can trigger the interests of HCI research community, to further explore how people's advanced cognitive constructions can inform the design and development of future empathic technologies.

\section{Related Work}
\label{related work}

The research on biodata representation has involved a variety of physiological signals, such as power of electromyography (pEMG), galvanic skin response (GSR), heart rate (HR), respiration effort (RE) and oxyhemoglobin saturation by pulse oximetry (SpO2) \cite{salminen_evoking_2019}. Research showed that GSR significantly enhanced physiological synchrony, and RE and HR provided the strongest user-perceived co-presence \cite{gong_comparing_2025}. In this work, we selected these three signals for investigating user's preferences for biodata representation.

Psychological models play a critical role in interpreting meaningful internal states from biodata. These models can be roughly divided into two categories: categorical models and dimensional models. 
Categorical approaches classify inner state into discrete categories (e.g., happiness, sadness, anger). Representative examples of this approach are Ekman’s six basic emotions \cite{ekman_universals_1971} and Plutchik’s emotional wheel model \cite{plutchik_emotions_2003}. 
By contrast, dimensional approaches map states onto continuous axes (e.g. valence-arousal), locating them within a shared affective space. Examples include the the Pleasure-Arousal-Dominance (PAD) model \cite{mehrabian1996pleasure} and the circumplex model of affect \cite{russell_circumplex_1980}. 
Among these, the circumplex model is currently the most widely used emotion classification framework and has been validated for its high accuracy \cite{halbig_systematic_2021, marin-morales_emotion_2020}. This model constructs a continuous, two-dimensional emotional space using valence (pleasure-displeasure) and arousal (activation-deactivation) dimensions, making it well-suited for mapping multisource physiological signals \cite{He_Zhou_Liu_Hu_2020}. Although there are also other internal states that can be implied from biodata, e.g. stress and attention, this study primarily investigates the emotion-related ones.

There is a plethora of research focused on multi-modal biodata representations and their impact on end user experience. For example, Dey et al. \cite{dey_effects_2018} found that perceiving another person's heartbeat through auditory and haptic can make participants more aware of the presence of others. Similarly, Salminen et al. \cite{salminen_evoking_2019} mapped respiration to aura to explore the relationship between respiration synchronizations and empathy. Furthermore, Lee et al. \cite{lee_understanding_2022} examined the influence of different visual representation styles on users' arousal levels, distraction and preferences. 
However, research on users’ preferences for expressing biodata and underlying cognitive systems remains limited. Consequently, determining a means of biodata representation aligned with the user’s cognition—particularly how the user understands, interprets and expresses biodata—remains challenging. Moreover, prior work typically examines biodata representation in isolated scenarios, leaving a critical gap in understanding how users conceptualize biosignals across diverse environments and generating contextual adaptive expression. These collective challenges motivate our exploration of GenAI as a potential pathway toward adaptive, human-cognitive-aligned biodata representation.

Over the last decade, we have witnessed a significant advance in AI technology. Notably, algorithms such as support vector machines (SVM) and K-nearest neighbors (KNN), have been proven effective in classifying and predicting emotional states using labeled biodata \cite{alharbi2024explainable}. The standard process typically involves several key steps: data acquisition, data preprocessing, feature extraction, model training and emotion classification. Many previous scholarly efforts have also been put into manual annotation of biosensory data \cite{petrescu_integrating_2020}, or human feedback on ML-based emotion classification \cite{mateos-garcia_driver_2023, yang_distinguishing_2019} etc. It is known as HITL, a collaborative approach that integrates human knowledge and expertise into the lifecycle of machine learning and AI systems. It is particularly valued by research community especially for human-related issues like emotional states, cognitive capabilities and medical fields that generally lack of training data \cite{Wu_Xiao_Sun_Zhang_Ma_He_2022}.  While borrowing from the concept of HITL, this study extracts human users' preferences and cognitive patterns for understanding and interpreting biosensory information, which is more abstract and implicit compared to verbal languages and visual images. This kind of human knowledge, sometimes even tacit and ambiguous, will then be fed to AIs and further guide them to generate biodata representations aligned with human cognitive preferences.

\section{User Elicitation Workshop}

To extract and understand users’ preferences regarding the expression of biodata in co-present virtual/hybrid scenarios, we conducted a user elicitation workshop with HCI experts through a structured protocol. We adopted this approach as it has already been widely applied to investigate users’ needs and preferences regarding their interactions with technical systems \cite{harrington2021eliciting, mulvale2019co}. 

\subsection{Workshop Setting}
To ensure the quality and relevance of feedback, we recruited 30 participants via social media platforms, all of whom either hold or are pursuing postgraduate degrees in HCI-related fields. The participants ranged in age from 21 to 50 years (Mean = 26.6, SD = 7.7), consisting of 18 males and 12 females. Regarding prior experience with virtual reality (VR), 2 participants (6.7\%) had never used VR, 14 participants (46.7\%) had less than one year of experience, 10 participants (33.3\%) had one to three years of experience, and 3 participants (10.0\%) had more than three years of experience. 

\begin{figure}[H]
  \centering
  \includegraphics[width=\linewidth]{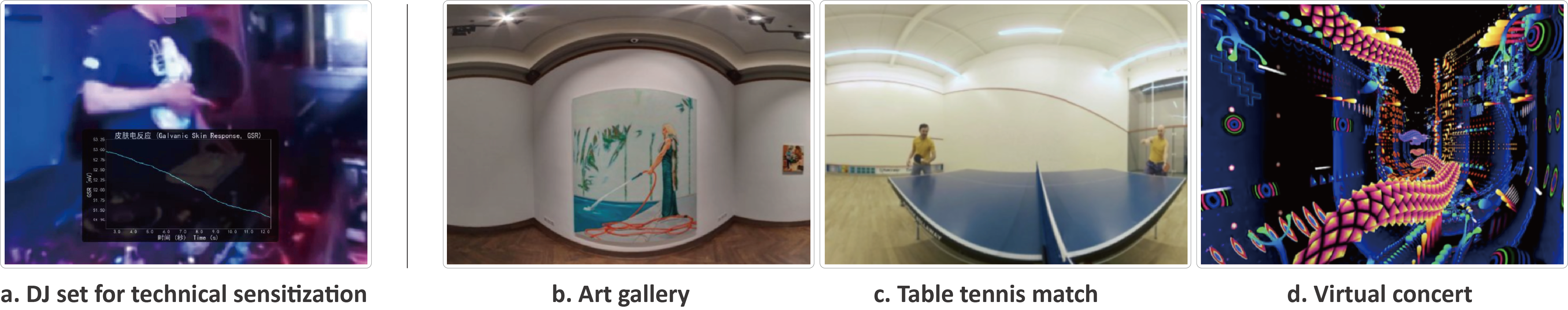}
  \caption{DJ set for technical sensitization and three representative co-present scenarios in VR.}
  \Description{Figure a shows the 360° DJ performance with overlaid biodata visualization used in the technology sensitization stage. Figures b, c, and d are the three typical co-present scenarios we selected: gallery, sports, and concert respectively.}
  \label{DJ set and three typical co-present VR scenarios}
\end{figure}

We provided each participant with one PICO 4 Pro VR headset to experience virtual scenes and one laptop computer with access to text-to-image GenAI tools (Midjourney, DALL-E and Stable Diffusion). To prepare the participants to ideate the biodata representations in the later session, the workshop began with a technical sensitization session. The participants were exposed to a 360-degree video clip of a DJ performance \footnote{\url{https://www.youtube.com/watch?v=DbUrKkb4JJ4}} with overlaid biodata information (Fig. \ref{DJ set and three typical co-present VR scenarios}a), to establish a shared understanding of how these signals enable the perception of "the co-present others". At this stage, plain biodata visualization was used to avoid biasing the participant's own representations. After the technical sensitization session, the participant then experienced three typical co-present VR scenarios in sequence: an art gallery, a table tennis match and a virtual concert (Fig. \ref{DJ set and three typical co-present VR scenarios}b, c, d), with each scenario lasting 50 seconds. Instead of depending solely on a particular type of biodata or a single scenario, we intentionally selected three representative target scenarios, each characterized by varying levels of co-present intensity and different user needs for experiencing co-presence. This allows us to concentrate on investigating the cognitive meta-constructions that are beyond and independent of specific scenarios or biosignals, while participants remained grounded in concrete contexts during ideation.

Then, the participant was asked to ideate on how to express biodata stemming from co-present others in all three target scenarios. We encouraged our participants to envision the overall scene using specific biosignals (GSR, RE, HR, see Section \ref{related work}) and to concretize their descriptions of the scene as possible. The participant then used these descriptions as text-to-image prompts for AI tools to generate the corresponding visual representations. Think Aloud strategy was employed during this process to better capture the participant's intention. The participant could try as many attempts as necessary, until they were satisfied with the AI-generated results and determined one image for each scenario to be uploaded along with the textual prompt via an online questionnaire. For each uploaded image, the participant would also need to select additional tags and ratings, or add supplementary information if necessary, for further elaborating what they had envisioned and created and why. This sort of "self-annotation" process helps clarify the participant's preferences and minimizes potential misunderstandings. 

Prior to the workshop, each participant was provided an informed consent form, which explained the research purpose and the use of the collected data. Participants were allowed to withdraw from the experiment at any time. All collected data were anonymized to protect the participants' privacy. Upon the completion of the experiment, each participant would receive a gift worth 25 RMB as compensation. This study was conducted in accordance with the ethical guidelines approved by the Institutional Review Board (IRB)  of the authors' affiliated institution.

\begin{figure}[h]
  \centering
  \includegraphics[width=0.6\linewidth]{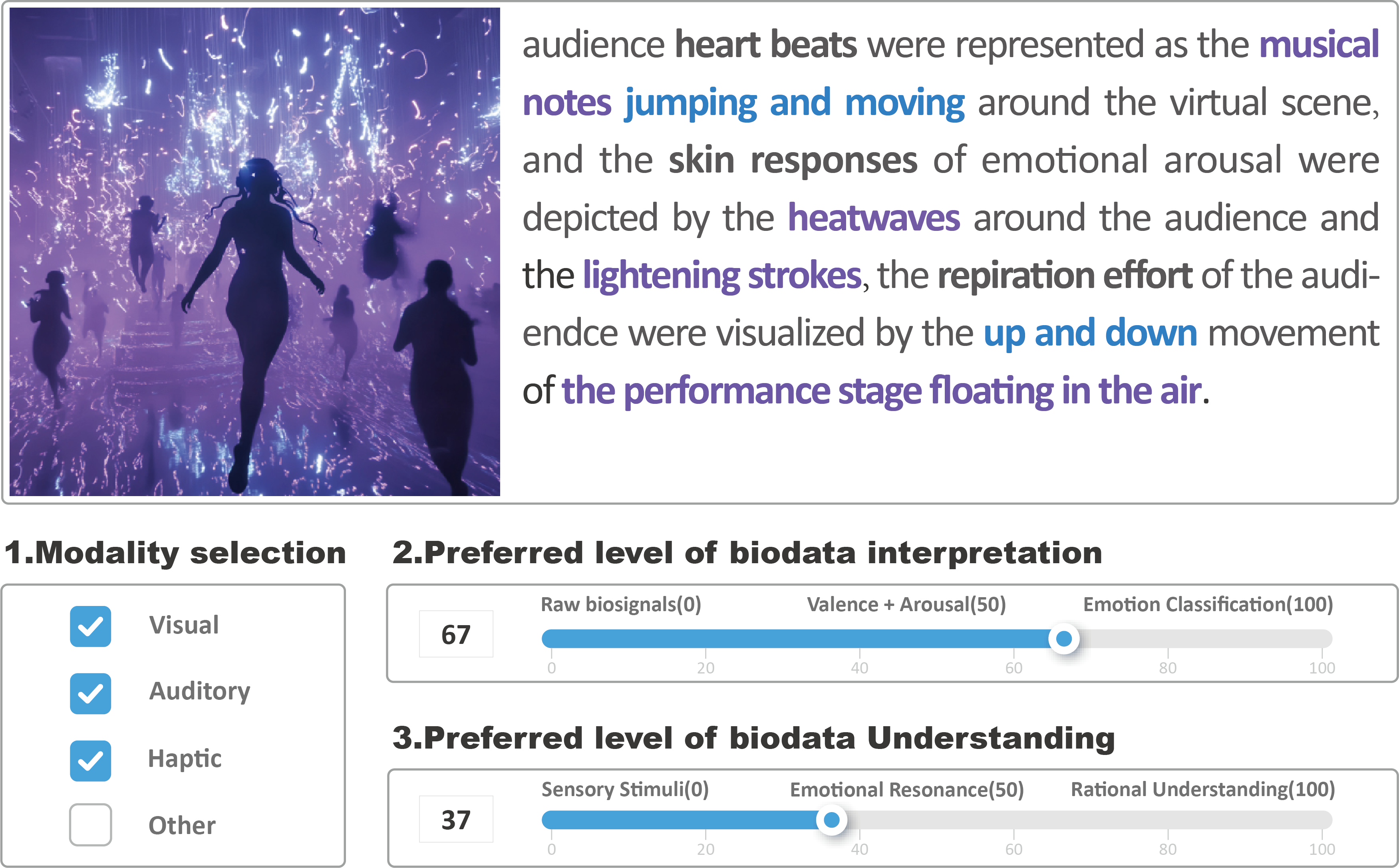}
  \caption{Key questionnaire results for P22.}
  \Description{This image shows the key questionnaire results of P22. The first part is the image he generated using the GenAI tool. The second part is his prompt'audience heart beats were represented as the musical notes jumping and moving around the virtual scene, and the skin responses of emotional arousa were depicted by the heatwaves around the audience and the lightening strokes, the respiration effort of the audiendce were visualized by the up and down movement of the performance stage floating in the air'. The third part is the modalities "Visual,Auditory,Haptic" that he uses when expressing biodata. The fourth and fifth parts are screenshots of the slider scale regarding preference for level of biodata interpretation and biodata understanding respectively, with values of 37 and 67}
  \label{P22}
\end{figure}

\subsection{Results}
\label{subsec: Results}

After a thorough scrutiny and team discussion of participants' responses, we decided to exclude certain participant feedback due to their overly simple nature and insufficient information for further analysis. As a result, we acquired a total of 71 questionnaire responses, respectively 24 for the virtual gallery scenario, 24 for the table tennis match and 23 for the virtual concert. To present an example, Fig. \ref{P22} shown a complete participant response from P22. The top listed the image that P22 generated for the virtual concert scenario and the textual prompt used, the detailed analysis of which will be discussed later at the end of this subsection. The bottom part of Fig. \ref{P22} showcased the three additional annotations that P22 attached to the above image, respectively: the \textbf{selected modality} for expressing biodata, the \textbf{preferred level of biodata interpretation} and the \textbf{preferred level of biodata understanding}. For the latter two, we provided two respective slider scales ranging from 0 to 100 for participants to further locate their creations. The leftmost end of the preferred level of biodata interpretation is labeled "raw biosignal", referring to unprocessed physiological signals without interpretation; While at the rightmost, "emotion classification" represents the accurate categorization of biodata into distinct emotional states. In between lies "valence + arousal", which indicates an intermediate level that maps biodata onto two continuous dimensions without precise categorization. All participants reported that they understood the scale as a continuous progression among the three levels. Similarly, the level of understanding reflects the extent to which the participant want their biodata representations to be comprehended. From left to right on the scale, the level of understanding gradually progresses through "sensory stimuli" and "emotional resonance" to "rational understanding". It is worth noting that as the level of understanding increases, the cognitive burden on users tends to increase as well. Next, we will provide a detailed account of the analytical processes we adopted for each of the aforementioned aspects and the results obtained.

\begin{figure}[h]
  \centering
  \includegraphics[width=\linewidth]{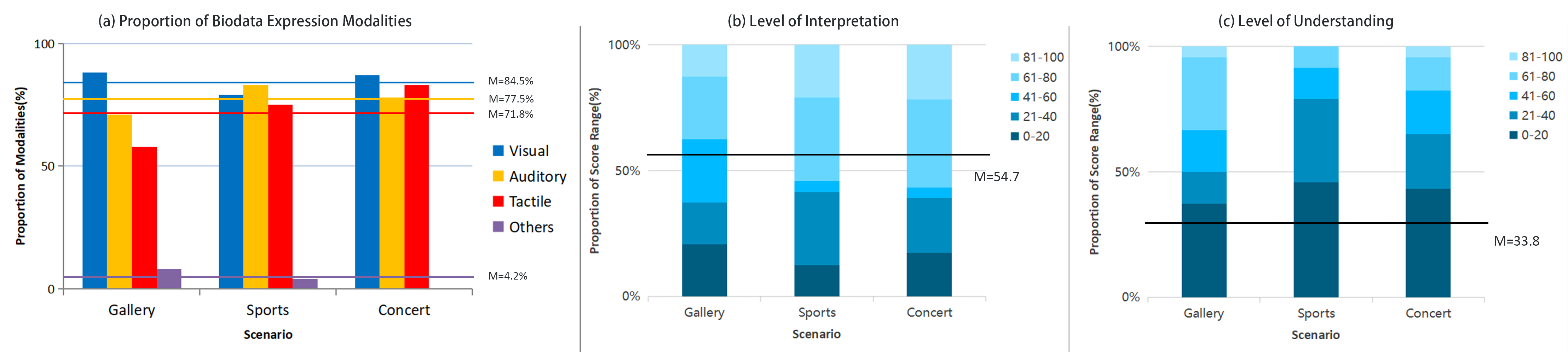}
  \caption{Data distribution of biodata expression modalities (a), mean scores for preferred level of Interpretation (b) and Understanding (c).}
  \Description{Mean scores for preferred information granularity (a), measured on a sliding scale from 0 (raw biosignals) to 50 (arousal+valence) to 100 (emotion classification), SD=26.9, 19.0, 27.2.
  Mean scores for prefered level of interpretability (b), measured on a sliding scale from 0 (sensory stimulation) to 50 (emotional resonance) to 100 (rational understanding), SD=25.4, 26.4, 30.7. Proportion of biodata expression modalities (c), mean proportion across three scenarios}
  \label{Data_Analysis}
\end{figure}

\textbf{Visual modality is the most commonly used for representing biodata among all modalities.} 
It is observed that similar to P22, most participants selected more than single modality. As shown in Fig. \ref{Data_Analysis}a, the proportion of visual, auditory and haptic modalities were 88\%, 71\% and 58\% for the gallery scenario; 79\%, 83\% and 75\% for the sports event; and 87\%, 78\% and 83\% for the virtual concert respectively. It can be implied that while multimodal biodata representations emerges as a common trend, the visual modality still remain dominant across different co-present scenarios. Moreover, the participants emphasized that the visual biodata representations must align with the context of co-present events. For example, P17 mentioned:"\textit{Virtual concerts are already rich in visual and audio elements, so I prefer to combine haptic feedback. The audience’s heart rate signals can be displayed in changing color hues using a smart wearable bracelet or fan props, which I can check occasionally.}" The limitations of human cognitive resources mean that we can only focus on a limited number of items at a time \cite{kahneman1973attention}. Therefore, our finding suggests that the use of visual biodata representations should not hinder the primary co-present experience. Instead, it should serve as a meaningful extension to it that makes you aware of the co-present others without actually distracting you, such as functioning as the frame or ambient lighting for a painting in a virtual gallery (P1, P6, P17, P22, P24, P25, P26), or as part of the stage effects at a concert (P8, P17, P11, P12, P18, P22, P23). Rather than plain data visualization, we can draw from the above user feedback that the design and dynamics of visual cues should be well situated within the co-present event context. 

\textbf{The participants prefer an overall moderate level of biodata interpretation.} Regarding the preferred level of biodata interpretation, P22 positioned his creation in the interval between "Valence + Arousal" (50) and "emotion classification" (100), leaning closer to the former (value = 67). We analyzed all the participants' responses across all target scenarios and summarized the results in Fig. \ref{Data_Analysis}b (\textit{M} = 54.75, \textit{SD} = 27.92). Compared to the other two scenarios, the participants' preferences for Gallery were relatively evenly distributed across all intervals, with slightly more data falling into [41, 60] and [61, 80]. In contrast, for Sports Game and Concert, we observed a tendency toward a bimodal distribution, where significantly less data fell within the [41, 60] interval, while more data were concentrated in [21, 40] and [61, 80]. It is important to clarify that, although we have presented the above data analysis and results, these data should not be considered as evaluative evidence, nor do they possess statistical significance due to the limited sample size. Instead, we view them more as an assistive tool for user elicitation research, aimed at revealing users' underlying cognitive frameworks by, for example, asking participants to further explain "WHY did you rate it this way?" As P22's comment on the gallery scenario:"\textit{All the biosignals could be designed to augment the experiences and it is not important in this context for audience to really distinguish among them. The atmosphere and vibes might be more significant than accurate representation and precise reading.}" An impressive consensus shared by multiple participants (P13, P17, P18, P22, P24, P26) is that in certain co-present experiences, people often do NOT need to rely on precise interpretations of the inner states of other individuals to perceive their presence. While completely unprocessed raw signals can sometimes lead to confusion about their meaning, a moderate level of ambiguity is entirely acceptable, and even encouraged. For instance, P17 and P24 both noted that in the context of sports events, they might prefer to understand the collective state of supporters on both sides during moments of victory or defeat, rather than precise physiological data about individual spectators.

\textbf{The participants prefer a lower understanding level between sensory stimuli to emotional resonance.} Regarding the level of biodata understanding, P22 marked his creation with the value 37, between sensory stimulation (0) and emotional resonance (50). As shown in Fig. \ref{Data_Analysis}c (\textit{M} = 33.82, \textit{SD} = 25.03), independent of whichever target scenario, we can observe a common and obvious trend: more than or at least half of the participants place their preferences within [0, 40], below the midpoint of emotional resonance; Meanwhile, the fewest users chose the interval [81, 100], indicating a marginal tendency toward rational understanding (100). The gallery is the only scenario with a relatively higher proportion of participants voting for emotional resonance and beyond. Several participants (P16, P17, P23, P24) explained that they were particularly interested in "\textit{understanding how others feel about and think of an piece of artwork}" and "\textit{resonating with different audiences over the same artwork}." In contrast, participants emphasized more on straightforward sensory stimuli with less cognitive exertion for the rest two scenarios: "\textit{I'm willing to open all my senses to immerse myself in a highly open and socially engaging atmosphere such as in a virtual concert.}" (P5) "\textit{Sports competitions are usually intensive and thus requires stronger sensory input.}" (P9) Compared to explicit social cues that are easy to capture and comprehend, e.g. verbal communication, facial expressions etc., biodata is more implicit and not yet familiar by the public. It thus imposes extra cognitive burdens on users to decode the information conveyed by this new technology-mediated medium. Presumably, it will take a longer period and more user training for such implicit social cues to be widely accepted in social contexts. For co-present experiences, biodata representations often need to provide the necessary complexity of social information in a form that is as intuitive and physically perceptible as possible.

\begin{figure}[h]
  \centering
  \includegraphics[width=\linewidth]{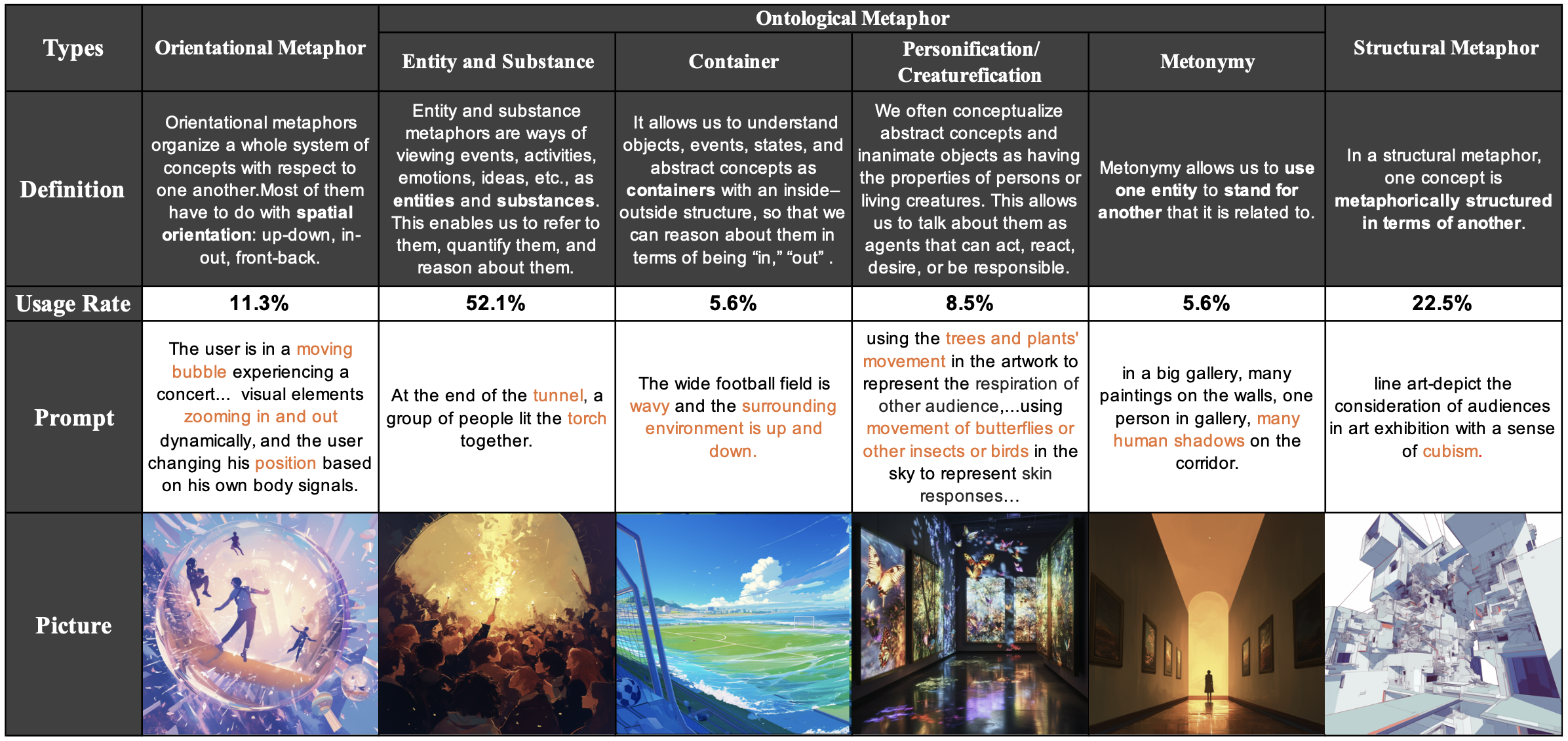}
  \caption{The classification, definition, usage proportion of metaphor, the corresponding prompts and pictures of the participants.}
  \Description{This picture shows the classification, definition usage proportion of metaphor, as well as the corresponding prompts and images of the participants.Metaphors are mainly divided into three categories, namely Orientational Metaphor, Ontological Metaphor and Structural Metaphor. Among them, Orientational Metaphor refers to express abstract concepts or things with specific spatial positions, like up-down, in-out front-behind. Among all the participants, 11.3\% of the pictures used such metaphor. The corresponding prompt for the participants is "The user is in a moving bubble experiencing a concert...  visual elements zooming in and out dynamically， and the user changing his position based on his own body signals". The Ontological Metaphor can be subdivided into four subcategories. The first category of Entity and Substance refers to Understand abstract concepts by conceptualizing them as objects, entities the corresponding participant prompts are "At the end of the tunnel, a group of people lit the torch together". Among all the participants, 52.1\% of the pictures used such metaphor. The second type of Container refers to understand abstract concepts by conceptualizing them as containers, corresponding "Understand abstract concepts by conceptualizing them as containers". Among all the participants, 5.6\% of the pictures used such metaphor. The third kind of Personification/Creaturefication is defined as "Personification or Creaturefication of the abstract thing", the corresponding prompt is "using the trees and plants' movement in the artwork to represent the respiration of other audience,... using movement of butterflies or other insects or birds in the sky to represent skin responses…”. Among all the participants, 8.5\% of the pictures used such metaphor. The last type is Metonymy. The definition is "Use one entity to stand for another that is closely related to it within the same cognitive domain. Among all the participants, 5.6\% of the pictures used such metaphor. The corresponding prompt is "in a big gallery, many paintings on the walls, one person in gallery, many human shadows on the corridor. The definition of structural metaphor is "Use a complete and familiar conceptual structure to organize or understand another more abstract conceptual structure. Among all the participants, 22.5\% of the pictures used such metaphor. The corresponding prompt is "line art-depict the consideration of audiences in art exhibition with a sense of cubism". Some participants used simple data visualization, touchability and audiodization. Such images do not fall into the category of metaphor, accounting for 28.2\%. Such pictures are presented in direct charts or other audible and touchable senses. The corresponding prompt is "Breathing and heart rate are converted into visual forms and music that are displayed in real time in the table tennis arena".}
  \label{fig:Metaphor Type}
\end{figure}

\textbf{Entity and substance metaphor is the most common vehicle to convey biosensory information.} We conducted a thematic analysis for analyzing participants' textual prompts. Three independent coders carried out open coding following an inductive reflexive approach suggested by \cite{williams2019art, braun2006using}. Specifically, we focused on coding 1) the types of imagery or methods participants used to express biodata, and 2) the associated dynamics of such expressions. Take P22's response as an example (Fig. \ref{P22}), he not only used natural phenomena such as "\textit{heatwaves}" and "\textit{lightening stokes}", but also leveraged spatial dynamics like "\textit{jumping and moving}" and "\textit{up and down}" to describe symbols ("\textit{musical notes}") and a physical object ("\textit{floating performance stage"}). The resulting codes mostly fall into nine types: Environmental information, living organisms, art styles, natural phenomena, human presence indicators, symbols, physical objects, multisensory effects and spatial dynamics. All the coding results were compared, and disagreements were resolved through discussions within the research team, ultimately achieving a consensus rate of over 99\%. In the following axial coding phase, we examined the relationships between the codes and clustered them into higher-level categories accordingly through affinity diagramming \cite{miles2014qualitative}. Our primary finding is that the majority of participant responses employed one or more metaphorical representations to express biodata. Given that earlier studies by Lackoff and others have already established systematic discourse on metaphors and related concepts \cite{lakoff2008metaphors}, we decided to draw upon their taxonomy to categorize our codes, and the results are summarized in Fig. \ref{fig:Metaphor Type}.

More than half of the participant responses (52.1\%) adopted \textbf{ontological metaphors}, explicitly the entity and substance metaphors, making it the most popular means to express biosensory information. The frequent use of "\textit{cloud}", "\textit{fog}" and "\textit{forest}" in the gallery scenario emphasized the role of biodata as ambient or peripheral information, minimizing interference with the main experience. Other ontological metaphors also include container metaphors, personification/creaturefication and metonymy, whose definitions were given in the Fig. \ref{fig:Metaphor Type}. This was followed by \textbf{structural metaphors} (22.5\%), which "use one highly structured and clearly delineated concept to structure another (usually more abstract concept)" \cite{lakoff2008metaphors}. A typical example by P3 was mapping the audience's inner state onto the art style of "\textit{cubism}". \textbf{Orientational metaphors} (11.3\%) appeared more frequently in the sports and concert scenarios, where a pair of opposing spatial indicators (e.g. "\textit{in and out}", "\textit{up and down}") were used to dynamically signify intensive shifts in psychophysiological states. In addition, 28.2\% of participant responses utilized plain data visualization/sonification/haptification either solely or in combination with metaphors. As in the case of P22, we can see an integrated use of entity and substance metaphors, orientational metaphors, and even structural metaphors, if we consider the analogy between biodata and musical notes can essentially be regarded as a mapping between two structural concepts. Such findings were both surprising and fascinating to us: when interpreting and communicating more abstract and less familiar information like biodata, people instinctively resort to metaphorical representations, by exploiting their shared experiential bases and pinpointing metaphors that are more straightforward to grasp thus more likely to gather empathy and understanding. It hence led us to propose our framework in the following section.

\section{BioMetaphor Framework}

\subsection{Construction of an GenAI-Driven Framework for Biodata Representation}

Our user elicitation workshop has highlighted \textit{metaphors} as a key cognitive mechanism users employ when expressing biodata. Metaphor is NOT merely linguistic; Instead, our findings strongly echoed Lackoff's statement: human thought processes are essentially metaphorical, and human conceptual system in itself is metaphorically structured and defined \cite{lakoff2008metaphors}. Informed by the previous workshop findings, we reached the following consensus on the overall design guidelines:
\begin{itemize}
    \item \textbf{Metaphor as the core mechanism}.
    \item \textbf{Visual representation as the primary modality}.
    \item \textbf{Moderate level of biodata interpretation}, meaning we opt for continuous, dimensional descriptors of inner states rather than raw physiological signals or categorical emotional information.
    \item \textbf{Low cognitive load representations}, favoring physically perceptible forms over cognitively demanding ones.
\end{itemize}
Based on these design guidelines, we propose a Gen-AI-driven framework for biodata representation, \textit{BioMetaphor}, which significantly distinguishes itself from prior studies. The framework comprises four compoments: 1) Biodata Modelling module, 2) LLM-driven Metaphor Building module, 3) Diffusion-based Representation Generation module, and 4) VR Integration module. Among these, the 2nd and the 3rd modules are the GenAI-powered core. Together, these components process biosensory information into metaphorical representations as visual social cues for co-present events. Fig.\ref{fig:teaser.png} details the components and their interactions. 

\subsubsection{Biodata Modelling Module}
This module decides the input signal for subsequent AI-driven inner state analysis. We drew upon Russell’s circumplex model of affect \cite{russell_circumplex_1980} to provide a moderate granularity of initial biosensory input, rather than starting from a precise emotion categorization. This model maps raw biodata to a pair of Valence-Arousal values (V-A pair). Valence indicates the degree of pleasantness, ranging continuously from negative (unpleasant) to positive (pleasant), while arousal measures the intensity of the emotion from low to high. For instance, a normalized V-A pair like (0.14, 0.85) locates an inner state in characterized by a relatively low valence value 0.14, signifying an unpleasant state, and a relatively high arousal value 0.85, indicating strong emotional intensity. Previous research has already proved the technical feasibility of mapping multi-source biosignals, e.g. HR, RE and GSR etc., to the according V-A pairs using machine learning \cite{mateos-garcia_driver_2023, marin2018affective}. However, particular biosignals and machine learning methods are considered out of our scope. We focus more on how the V-A pairs, as an initial input that reflects an intermediate level of inner states, can be further understood and interpreted by current GenAI. Therefore, instead of using real biodata, we used a simulated dataset of normalized V-A pairs, which derived from the polar transformations of prototypical emotion coordinates distributed within the circumplex model. This choice helps eliminate sensor noise and inter-subject variability, which enables core logic validation and cross-study compatibility \cite{lanovaz2020machine}. To note that, although we used the circumplex model, the framework per se does not reject other dimensional affective models, such as the 3D emotion space model which maps people's inner states onto a three-dimensional, continuous coordinate system.

\begin{figure}[h]
  \centering
  \includegraphics[width=\linewidth]{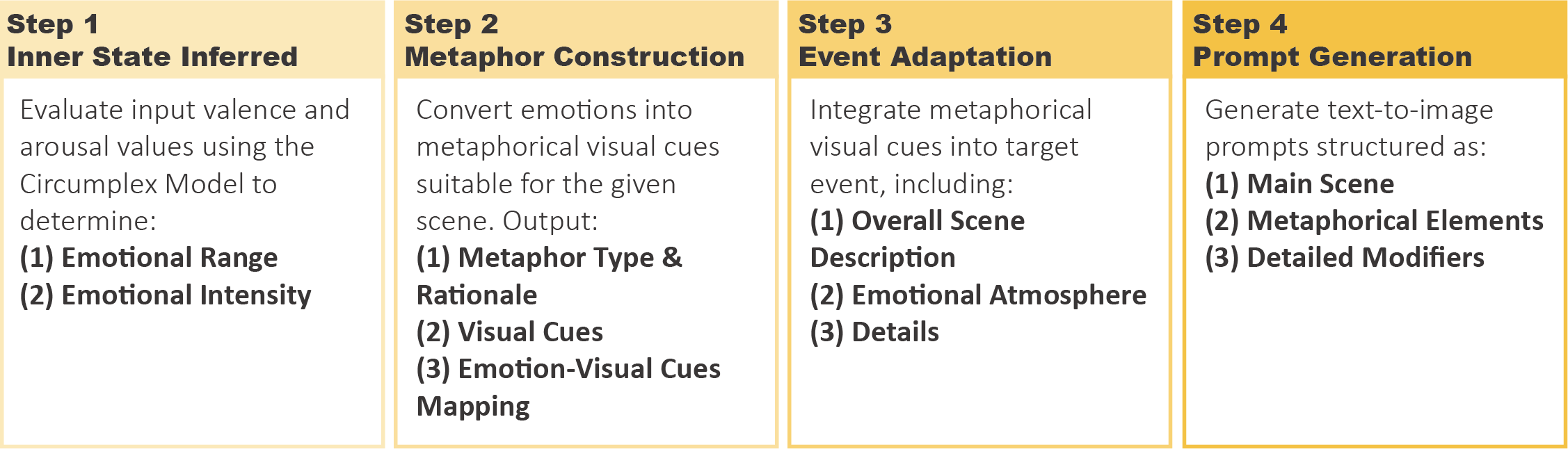}
  \caption{A  four-step Chain-of-Thought process.}
  \Description{The image is a diagram illustrating a four-step process for CoT prompt engineering using Large Language Models (LLMs). The diagram is divided into four vertical sections, each representing a step in the process.
  Step 1: INNER STATE INFERRED
  Description: Evaluate input valence and arousal values using the Valence-Arousal Model to determine:1) Emotional Range; 2) Emotional Intensity.
  Step 2: METAPHOR CONSTRUCTION
  Description: Convert emotions into metaphorical visual cues suitable for the given scene. Output: 1) Metaphor Type and Rationale; 2) Visual Cues; 3) Emotion-Visual Cues Mapping.
  Step 3: EVENT ADAPTATION
  Description: Integrate metaphorical visual cues into target event, including: 1) Overall Scene Description; 2) Emotional Atmosphere; 3) Details.
  Step 4: PROMPT GENERATION
  Description: Generate text-to-image prompts structured as:
  1) Main Scene; 2) Metaphorical Elements; 3) Detailed Modifiers.}
  \label{fig:CoT_Steps}
\end{figure}

\subsubsection{Metaphor Building Module}
This module addresses the translation of V-A pairs into context-specific metaphorical descriptions using LLMs. It simulates the users’ cognitive process of metaphor construction by inferring inner state and expressing it metaphorically, finally generating prompts that contain metaphorical visual cues. To achieve this with structured reasoning, we employ Chain-of-Thought (CoT) prompting \cite{wei2022chain}, a technique that guides the LLMs to break down complex tasks into intermediate reasoning steps, enabling the model to understand and execute the operations required at each stage of reasoning. By organizing a clear, step-by-step process from biodata interpretation to metaphorical expression, CoT enables the generation process of the final textual prompt to be visible to the researchers, allowing us to trace and control any intermediate steps. As shown in Fig. \ref{fig:CoT_Steps}, the CoT procedure consists of four steps: 

\textbf{Inner State Inference.} The CoT prompting begins with inferring the emotional range from the biodata. Specifically, the LLM performs a data analysis on the input V-A pair based on the circumplex model, deriving the corresponding inner states. This step is designed to output emotional range and emotional intensity for subsequent metaphor construction. When we refer to "inner states", it includes, but is not limited to, emotion-related states. We chose emotions because they are pervasive; however, other meaningful inner states, such as attention and stress, can also be extracted from biodata and are therefore applicable within this framework as well.

\textbf{Metaphor Construction.} This step involves mapping the inner states obtained from the previous step into metaphorical visual cues. We provide definitions and examples of three metaphor types derived from workshops—\textit{orientational metaphor, ontological metaphor} and \textit{structural metaphor}. The LLM is instructed to select from these predefined metaphor types and generate visual cues based on contextual characteristics. We require it to output the chosen metaphor type and the rationale, the corresponding visual cues, and the mapping between inner states and visual cues, while impose no preference for any specific metaphor type.

\textbf{Event Adaptation.} The metaphorical visual cues need to be further contextualized and situated appropriately within the target co-present events accordingly. To this end, this step instructs the LLM to combine the visual cues with the target scene and produce a comprehensive description of the scene. The description should include: 1) an overall scene description, 2) the emotional atmosphere, and 3) more detailed elements of the scene. The prompt engineering also ensures that metaphorical visual cues do not affect the main activity when combined into the event. 
%For example, in a gallery scene, the appearance of metaphorical visual cues should not hide the painting and affect the viewing experience.

\textbf{Prompt Generation.} In the last step, the event-adapted metaphors are converted into structured natural language descriptions. Compatibility with mainstream text-to-image models is ensured by instructing the LLM to organize and output the textual prompt according to a three-layer structure: 1) Main Scene, 2) Metaphorical Elements, and 3) Detailed Modifiers (e.g., the layering of light, the contrast of texture), which serves as input for the following module. 

\subsubsection{Representation Generation Module}
This module transforms the metaphor-rich textual prompts generated previously by LLM, to panoramic VR-ready images. In this study, we used Stable Diffusion (SD) to implement text-to-image generation. However, the framework itself does not restrict users from using other GenAI models. We further combined SDXL along with Low-Rank Adaptation (LoRA) to enable the rendering and visualization of 360-degree static panoramic scenes. In practical implementation, risks such as image inconsistency may arise, potentially disrupting narrative coherence or breaking user immersion. Thus, we recommend several techniques such as fixing the random seed \cite{manas2024improving}, using image-to-image \cite{isola2017image} to reference an initial image, or incorporating users to select a preferred depiction as an anchor for subsequent generations.

\subsubsection{Virtual Integration Module}

This module takes care of merging the generated metaphorical biodata representations with the actual activity scenarios and loading the results onto the user's output devices, where the final end user experience is shaped and delivered. Currently, we have realized this process using Unity 3D Engine. The Base64-encoded image data are decoded into byte arrays (binary image data) to create new Texture 2D material, which is then projected onto the skybox. The final 3D scene was then streamed to the user's head-mounted display (HMD), e.g. PICO 4 Pro in our implementation, offering a spherical immersive experience. It is important to clarify that the proposed framework is not confined to the specific instantiation we provided, nor is it solely applicable to fully virtual co-present events. For instance, in the case of hybrid art exhibitions, we can follow the same approach by using projection mapping to display the visual representations generated from online or offline audiences' biodata, onto or around the physical artworks. It may not only extend the artistic expression of the physical pieces, but also serve as a means to enrich the on-site viewing experience. 

In conclusion, the proposed framework does not rely exclusively on any specific technical implementation. Rather, it aims to provide a high-level approach that is flexible enough to incorporate GenAI-driven biodata representations into various co-present experiences, while also accommodating a wide range of situational requirements and technical combinations.

\begin{figure}[H]
  \centering
  \includegraphics[width=\linewidth]{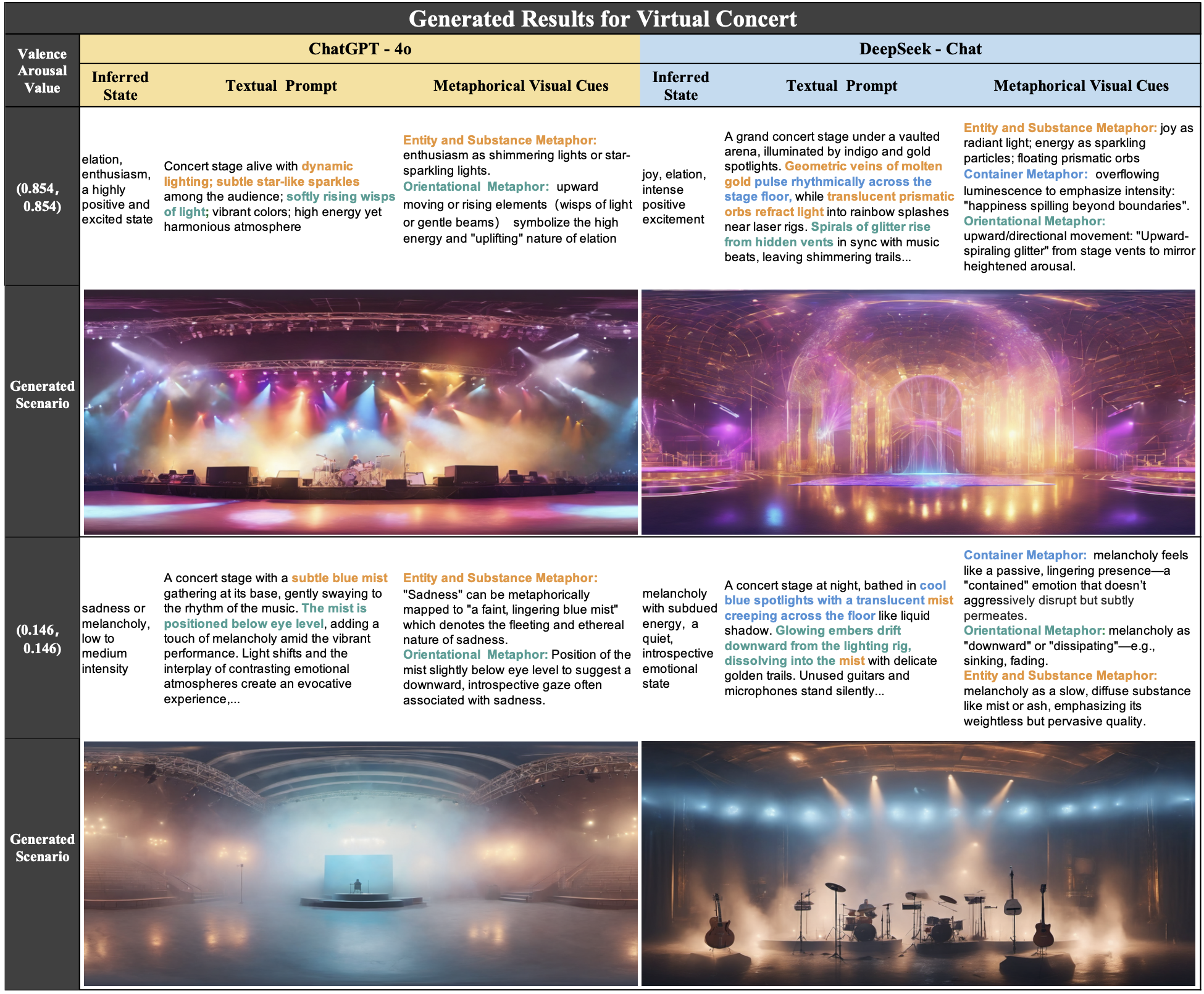}
  \caption{Examples of Results Generated by the BioMetaphor Framework}
  \Description{The figure is a table presents a detailed comparison of generated results for virtual concert scenarios using two large language models, GPT-4o and DeepSeek-Chat, as part of the BioMetaphor framework's results showcase. It is structured as a table with two main columns. The left column for GPT-4o and the right column for DeepSeek-Chat. Each column includes subcategories labeled "Valence Arousal Value," "Inferred State," "Textual Prompt," "Metaphor Types and Visual Cues," and "Generated Scenario." Each column contains text descriptions and an accompanying image for the generated scenario, facilitating a contrast between the outputs of the two models. 

  The left column is the GPT-4o Section. When the input valence arousal value is (0.854, 0.854), the inferred state is: elation, enthusiasm, a highly positive and excited state; the generated Metaphor Types and Visual Cues are: Entity and Substance Metaphor: enthusiasm as shimmering lights or sparkling lights; Oriental Metaphor: upward moving or rising elements (wisps of light or gentle beams), symbolize the high energy and uplifting nature of elation. The corresponding textual prompt is: "Concert stage alive with dynamic lighting; subtle star-like sparkles among the audience; softly rising wisps of light, vibrant colors; high energy yet harmonious atmosphere." And the generated scenario image based on the textual prompt depicts a vibrant concert stage illuminated with dynamic lighting. The scene features colorful beams of light, primarily in shades of purple, pink, and blue, creating a high-energy atmosphere. Subtle star-like sparkles are visible among the audience, enhancing the sense of elation and excitement. The stage appears to be in an open-air setting with a crowd of people gathered, and the lighting effects contribute to a harmonious yet lively ambiance.

  When the input valence arousal value is (0.146, 0.146), the inferred state is: sadness or melancholy, low to medium intensity; the generated Metaphor Types and Visual Cues are:  Entity and Substance Metaphor： "Sadness" can be  metaphorically mapped to "a faint, lingering blue mist" which denotes the fleeting and ethereal nature of sadness. Orientational  Metaphor: Position of the mist slightly below eye level to suggest a downward, introspective gaze often associated with sadness.  The corresponding textual prompt is: "A concert stage with a subtle blue mist gathering at its base, gently swaying to the rhythm of the music. The mist is positioned below eye level, adding a touch of melancholy amid the vibrant performance. Light shifts and the interplay of contrasting emotional atmospheres create an evocative experience,..." And the generated scenario image The image portrays a concert stage enveloped in a subtle blue mist, set within an arched architectural structure with tall columns.   The mist gathers at the base and on the stage air, showing a melancholic and introspective mood. The lighting is soft and muted, with a hazy blue tone dominating the scene.
  
  The right column is the DeepSeek-Chat Section. When the input valence arousal value is (0.854, 0.854), the inferred state is: joy, elation, intense positive excitement; the generated Metaphor Types and Visual cues are: Entity and Substance Metaphor: joy as radiant light; energy as sparkling particles; floating prismatic orbs. Container Metaphor：overflowing luminescence to emphasize intensity: "happiness spilling beyond boundaries". Orientational Metaphor: upward/directional movement: "Upward-spiraling glitter" from stage vents to mirror heightened arousal.. The corresponding textual prompt is: "Concert stage alive with dynamic lighting; subtle star-like sparkles among the audience; softly rising wisps of light, vibrant colors; high energy yet harmonious atmosphere." And the generated scenario image based on the textual prompt shows a concert stage at night, bathed in the golden spotlights with veins of molten gold creeping across the floor like liquid shadow. The stage features an arched backdrop with delicate architectural elements and a cool blue light in the middle of the stage，and glowing embers drift downward from the lighting rig, dissolving into the prismatic obs floor with delicate golden trails. The overall color of subtle golden accents dominates the scene.

   When the input valence arousal value is (0.146, 0.146), the inferred state is: melancholy with subdued energy,  a quiet, introspective emotional state; the generated Metaphor Types and Visual Cues are: Container Metaphor:  melancholy feels like a passive, lingering presence—a "contained" emotion that doesn’t aggressively disrupt but subtly permeates.  Orientational Metaphor: melancholy as "downward" or "dissipating"—e.g., sinking, fading. Entity and Substance Metaphor: melancholy as a slow, diffuse substance like mist or ash, emphasizing its weightless but pervasive quality. The corresponding textual prompt is: "A concert stage at night, bathed in cool blue spotlights with a translucent mist creeping across the floor like liquid shadow. Glowing embers drift downward from the lighting rig, dissolving into the mist with delicate golden trails. Unused guitars and microphones stand silently..." And the generated scenario image based on the textual prompt depicts a concert stage covered in misty haze, with beams of light in orange, pink, and blue cutting through, creating a dramatic vibe. It features a drum kit, microphone stands, speakers, amps, and a keyboard setup. The reflective floor adds an ethereal touch.
}
  \label{fig:Framework_Result}
\end{figure}

\subsection{Framework Demonstration}

This section demonstrates one possible, but not the only, instantiation of the proposed \textit{BioMetaphor} framework. It is important to emphasize that our goal is not to devise a ready-to-use GenAI system or tool; rather, by building a minimal working prototype, we aim to rapidly validate the feasibility and generalizability of our proposed framework independent of particular technologies and implementations. Specifically, we seek to investigate: 1) whether current GenAI can properly interpret and reason about people’s inner states based on biosensory information with a certain degree of ambiguity; 2) whether current GenAI can learn from people about their preferences and tendencies in communicating biodata, and further imitate advanced cognitive mechanisms such as metaphorical representations; and 3) whether current GenAI can appropriately map inferred inner states onto metaphorical representations to generate visual social cues that align with human needs and event contexts.

To compare and analyze the complete reasoning chains over different models, we adopted two mainstream LLMs: GPT-4o and DeepSeek-Chat, for the construction of metaphorical visual cues. To avoid additional biases from text-to-image models that are unrelated to reasoning per se, we intentionally used only SDXL for the representation generation. Both GPT-4o and DeepSeek-Chat adopted the official recommended temperature for general conversation (GPT-4o = 1, DeepSeek-Chat = 1.3). 
% comparing the performance of two LLMs in reasoning about inner states, learning metaphorical thinking, and mapping inner states to visual cues.
In total, we tested eight prototypical V-A pairs, evenly distributed at 45-degree intervals in the circumplex model. These pairs were tested across all three co-present scenarios (gallery, sports, concert), resulting in a sum of 48 outcomes (24 outcomes per LLM). Fig. \ref{fig:Framework_Result} depicts four sample results, showcasing the corresponding inferred state, textual prompt, metaphorical visual cues and generated images. These results were both from the concert scenario, but corresponded to two contrasting VA pairs: 1) high arousal and valence (0.854, 0.854); and 2) low arousal and valence (0.146, 0.146).

Firstly, by comparing AI-inferred emotional ranges with the circumplex model as ground truth, we found that both LLMs analyzed and reasoned about the input VA pairs with basic accuracy, and no significantly contradictory outliers were observed. Secondly, both LLMs demonstrated the ability to flexibly combine more than one type of metaphor to express emotional states in a style approximating that of human users. Moreover, the selection and composition specific visual cues, e.g. "\textit{upward motion corresponding to lightness and joy, downward motion corresponding to heaviness and sadness}", or "\textit{cool tones corresponding to calmness and warm tones to liveliness}", both reflect that current GenAIs are, to some extent, capable of correctly grasping the characteristics of metaphorical expression.

However, we also observed that for structural metaphors, a more systematic and high-level metaphor type, the textual prompts generated by both LLMs exhibited varying degrees of oversimplification and superficiality. This indicates that current GenAIs still fail to capture the multifacetedness and semantic nuance involved in mapping between two complex conceptual domains. Additional supervision and control may therefore be required to ensure the quality of AI-generated metaphors, particularly in terms of whether these metaphors effectively convey users’ intended meanings, and whether they align with human social and cultural norms.

Last but not least, both LLMs were able to consider and respond to specific contextual requirements based on prompt engineering instructions, such as “\textit{‘Black dust’ is designed to seamlessly integrate into the gallery space without disrupting its primary activities (G4)}” or “\textit{‘Bright, shimmering lights’ subtly integrate with a concert scene, enhancing the atmosphere without overwhelming the primary focus(C1).}” However, upon examining the final image outputs, we observed that certain highly detailed textual prompts could not be fully realized by SDXL, resulting in the noticeable absence of metaphorical visual cues. We believe this reflects the current limitations of text-to-image AIs, requiring users to balance the GenAI capabilities with the detailedness of textual prompts in actual practice. Nevertheless, this issue is considered out of this study's scope. The complete demonstration samples can be referred to in the supplementary materials we submitted.

\section{Discussion}

\subsection{RQ1: User's Preference of Biodata Representations in Co-present Events}

One of the key findings from our user elicitation workshop was people's preference for using metaphorical visual cues to express physiological signals. Existing cognitive research has found that vision is the primary input channel for human sensory information, accounting for approximately 70\% of information processing capacity \cite{bruce2014visual}. To avoid cognitive overload for users, we suggest making effective use of users' peripheral vision. Furthermore, we recommend that biodata visual cues be backgrounded by subtly integrating them with ambient elements, such as lighting, vegetation and other environmental features. While this study primarily focused on visual metaphors, we hope to spark greater interest in the HCI community to systematically explore other modalities, such as audio and haptic metaphors, for multimodal biodata representations. Building upon this work, we believe that future research in this direction will have positive and lasting impacts on next-generation virtual and hybrid applications.

Our user study revealed a strong contextual dependency in participants' preferences for interpreting and understanding biodata. In low-paced and less intensive co-present settings like art galleries, users exhibited more diverse preferences for the granularity of biosensory information, and significantly more participants were willing to rationally understand others' biodata. In contrast, in dynamic and highly intensive co-present settings like sports games and concerts, while participants' preferences for biodata interpretation clustered around the moderate level in a bimodal distribution, their preferred level of understanding leaned more towards sensory stimulation and slightly below emotional resonance. According to Dual-Process Theory \cite{evans2013dual}, human thought processes are shaped by the combined influence of the fast, intuitive System 1, and the slow, analytical System 2. The implication of this theory for virtual or hybrid co-present applications is that to perceive or feel the "presence" of others, we often do not need very precise and detailed information, nor do we need to fully process or comprehend it. Simply adding more visual or auditory stimuli or increasing the complexity of information does not necessarily contribute to an enhanced sense of co-presence. Moreover, for scenarios where users' biodata tend to stay static or change slowly, more attention should be given to whether the stimulus intensity is sufficient to make the biodata cues perceptible.

Meanwhile, an additional potential explanation for the relatively discrete data distribution lies in the inherent diversity of individuals and their cognitive variability. For example, influenced by different cognitive styles, people with high need for cognition (NfC) prefer engaging in analytical reasoning, while those with high need for affect (NfA) or heightened sensitivity are more likely to prefer sensory stimulation \cite{Haddock_Maio_Arnold_Huskinson_2008}. If there is a clear target user group, it is strongly recommended to conduct prior cognitive profiling to inform the design of biodata representations. If the target user group is unknown or a wider public is involved, we can either utilize attention tracking technology to monitor the users' "focus loci" (points of attention) in real time to dynamically determine whether to provide fine-grained biosensory information, or offer customizable biodata representations. In this field, future quantitative research and controlled experiments may enhance the explainability of phenomena and deepen our understanding of causality. However, we must emphasize that the co-present experience itself is a complex continuum characterized by the interplay of multiple cognitive, emotional and behavioral processes. The significance of this study lies in extracting meaningful and actionable implications for designing and constructing a holistic co-present experience.

\subsection{RQ2: GenAI for Biodata Representations in Co-present Events}

Lake et al. used to compare and summarize the performance of artificial intelligence and humans in learning to complete specific tasks in \cite{lake2017building}. They pointed out that current AI lacks certain advanced cognitive abilities inherent to humans, such as compositional thinking about part-whole relationships, and causal reasoning about phenomena and their underlying causes. One of the most significant differences between machine learning and human learning is that machine learning largely relies on inductively learning latent patterns from massive amounts of data. In contrast, humans not only learn inductively, but also deductively and abductively. We argue that metaphor is a pervasive cognitive process of understanding a more abstract or complex concept through a simpler, more familiar one. Thus, it explains why, when dealing with biodata as a sort of implicit and less familiar social cues, people tend to intuitively adopt familiar imagery from everyday life to establish metaphorical expressions. As you may already notice, in virtual or hybrid co-present events, our attempts to use physiological signals to replace more commonly used methods of communication—such as verbal speech, facial expressions, or body posture—are essentially a form of metaphor: biodata is another language.

There is a growing body of research aimed at integrating human cognitive mechanisms for learning, reasoning and problem-solving into AI workflows. For example, Park et al. modeled long-term memory and reflection mechanisms in humans to construct smart agents \cite{park2023generative}. Other examples include \cite{sumers2023cognitive}, \cite{qu2024integration} and so forth. In our study, we attempted to use CoT prompting techniques to enable GenAI to express biodata through metaphorical thinking. Our work has preliminarily demonstrated that mainstream GenAIs are generally capable of learning and imitating this method. However, when it comes to more advanced concepts (e.g., structural metaphor), their understanding and application still lack a certain level of "depth", a limitation that is also observed in other domains. We insist that metaphor, as a universal and sophisticated cognitive mechanism, reflects some of the most fundamental aspects of human thinking. The incorporation of metaphor can help AI build a conceptual framework more consistent with human cognition, thereby enhancing its applicability to scenarios including, but not limited to, co-present experiences. At the same time, it is crucial to recognize that metaphors often carry rich and sometimes subtle social and cultural implications. We must carefully examine the appropriateness of AI-generated metaphors to prevent their potential misuse or misinterpretation.

\section{Conclusion}

This paper explores two RQs: 1) What do people prefer for expressing biodata in co-present events? 2) How can GenAI be used to generate biodata representations that align with human preferences? To answer the RQs, we first conducted a user elicitation workshop with 30 HCI experts to explore their preferences for biodata expression. Building on the resulting insights, we present our framework \textit{BioMetaphor} and its demonstration, to showcase the use of AI-generated metaphorical visual cues in different co-present event contexts.

We are aware of several limitations of the current study. On one hand, the proposed framework is a general architecture, rather than a ready-to-use system or AI tool. Therefore, future work may need to develop instantiations of this framework based on a concrete application scenario, allowing for a comprehensive and solid user validation and testing the framework's practical effects, such as whether it can indeed generate meaningful biodata representations that align with the consensus of most users. On the other hand, although we have provided a preliminary framework demonstration, it heavily relies on the quality of prompt engineering. Future studies may need to compare different prompt engineering strategies or incorporate fine-tuning methods beyond prompt engineering to further establish best practices for the framework. 

At the same time, this work makes several key contributions. First, through a qualitative analysis of user elicitation workshop results, we uncovered human preferences for interpreting and expressing biodata in co-present events. Second, we introduced \textit{BioMetaphor}, a GenAI-driven framework that expresses biodata in a way that reflects human cognitive preferences. Third, we demonstrated the capability of current GenAI systems to learn and partially imitate advanced human cognitive patterns. Moreover, we believe that the overall HITL-style approach prompts new possibilities of conducting human-inspired AI research. It is not confined to the current target application and is readily transferable to other domains beyond biodata representation. We hope this research sparks the HCI community's interest to further investigate how human cognitive constructions can foster more adaptive and empathic technologies.

% \begin{acks}
%   This work employs GPT-4o and DeepSeek-Chat for framework validation via prompt-driven semantic translation and Stable Diffusion XL for image generation, with representative outputs documented in Fig. \ref{fig:Framework_Result}. Regarding the manuscript, all contents are originally written by the author team, and GenAI tools were only responsible for refining the grammar and wording.
% \end{acks}

\bibliographystyle{ACM-Reference-Format}
\bibliography{sample-base}


\begin{thebibliography}{49}

%%% the Web address.

\ifx \showCODEN    \undefined \def \showCODEN     #1{\unskip}     \fi
\ifx \showISBNx    \undefined \def \showISBNx     #1{\unskip}     \fi
\ifx \showISBNxiii \undefined \def \showISBNxiii  #1{\unskip}     \fi
\ifx \showISSN     \undefined \def \showISSN      #1{\unskip}     \fi
\ifx \showLCCN     \undefined \def \showLCCN      #1{\unskip}     \fi
\ifx \shownote     \undefined \def \shownote      #1{#1}          \fi
\ifx \showarticletitle \undefined \def \showarticletitle #1{#1}   \fi
\ifx \showURL      \undefined \def \showURL       {\relax}        \fi
% invisible to TeX
\providecommand\bibfield[2]{#2}
\providecommand\bibinfo[2]{#2}
\providecommand\natexlab[1]{#1}
\providecommand\showeprint[2][]{arXiv:#2}

\bibitem[Alharbi(2024)]%
        {alharbi2024explainable}
\bibfield{author}{\bibinfo{person}{Hadeel Alharbi}.} \bibinfo{year}{2024}\natexlab{}.
\newblock \showarticletitle{Explainable feature selection and deep learning based emotion recognition in virtual reality using eye tracker and physiological data}.
\newblock \bibinfo{journal}{\emph{Frontiers in Medicine}}  \bibinfo{volume}{11} (\bibinfo{year}{2024}), \bibinfo{pages}{1438720}.
\newblock


\bibitem[Andrews(2020)]%
        {andrews2020concerts}
\bibfield{author}{\bibinfo{person}{Travis~M Andrews}.} \bibinfo{year}{2020}\natexlab{}.
\newblock \showarticletitle{Concerts are canceled, so Coldplay, John Legend and Keith Urban are playing right in your living rooms}.
\newblock \bibinfo{journal}{\emph{The Washington Post}} (\bibinfo{year}{2020}).
\newblock


\bibitem[Bender et~al\mbox{.}(2021)]%
        {Bender_Gebru_McMillan-Major_Shmitchell_2021}
\bibfield{author}{\bibinfo{person}{Emily~M. Bender}, \bibinfo{person}{Timnit Gebru}, \bibinfo{person}{Angelina McMillan-Major}, {and} \bibinfo{person}{Shmargaret Shmitchell}.} \bibinfo{year}{2021}\natexlab{}.
\newblock \showarticletitle{On the Dangers of Stochastic Parrots: Can Language Models Be Too Big?}. In \bibinfo{booktitle}{\emph{Proceedings of the 2021 ACM Conference on Fairness, Accountability, and Transparency}} \emph{(\bibinfo{series}{FAccT ’21})}. \bibinfo{publisher}{Association for Computing Machinery}, \bibinfo{address}{New York, NY, USA}, \bibinfo{pages}{610–623}.
\newblock
\showISBNx{978-1-4503-8309-7}
\href{https://doi.org/10.1145/3442188.3445922}{doi:\nolinkurl{10.1145/3442188.3445922}}


\bibitem[Braun and Clarke(2006)]%
        {braun2006using}
\bibfield{author}{\bibinfo{person}{Virginia Braun} {and} \bibinfo{person}{Victoria Clarke}.} \bibinfo{year}{2006}\natexlab{}.
\newblock \showarticletitle{Using thematic analysis in psychology}.
\newblock \bibinfo{journal}{\emph{Qualitative research in psychology}} \bibinfo{volume}{3}, \bibinfo{number}{2} (\bibinfo{year}{2006}), \bibinfo{pages}{77--101}.
\newblock


\bibitem[Bruce et~al\mbox{.}(2014)]%
        {bruce2014visual}
\bibfield{author}{\bibinfo{person}{Vicki Bruce}, \bibinfo{person}{Mark~A Georgeson}, {and} \bibinfo{person}{Patrick~R Green}.} \bibinfo{year}{2014}\natexlab{}.
\newblock \bibinfo{booktitle}{\emph{Visual perception: Physiology, psychology and ecology}}.
\newblock \bibinfo{publisher}{Psychology Press}.
\newblock


\bibitem[Curran et~al\mbox{.}(2019)]%
        {curran_understanding_2019}
\bibfield{author}{\bibinfo{person}{Max~T. Curran}, \bibinfo{person}{{View Profile}}, \bibinfo{person}{Jeremy~Raboff Gordon}, \bibinfo{person}{{View Profile}}, \bibinfo{person}{Lily Lin}, \bibinfo{person}{{View Profile}}, \bibinfo{person}{Priyashri~Kamlesh Sridhar}, \bibinfo{person}{{View Profile}}, \bibinfo{person}{John Chuang}, {and} \bibinfo{person}{{View Profile}}.} \bibinfo{year}{2019}\natexlab{}.
\newblock \showarticletitle{Understanding {Digitally}-{Mediated} {Empathy}}.
\newblock \bibinfo{journal}{\emph{Proceedings of the 2019 CHI Conference on Human Factors in Computing Systems}} (\bibinfo{date}{May} \bibinfo{year}{2019}), \bibinfo{pages}{1--13}.
\newblock
\showISSN{9781450359702}
\href{https://doi.org/10.1145/3290605.3300844}{doi:\nolinkurl{10.1145/3290605.3300844}}


\bibitem[Dautenhahn(1998)]%
        {Dautenhahn_1998}
\bibfield{author}{\bibinfo{person}{Kerstin Dautenhahn}.} \bibinfo{year}{1998}\natexlab{}.
\newblock \showarticletitle{THE ART OF DESIGNING SOCIALLY INTELLIGENT AGENTS: SCIENCE, FICTION, AND THE HUMAN IN THE LOOP}.
\newblock \bibinfo{journal}{\emph{Applied Artificial Intelligence}} \bibinfo{volume}{12}, \bibinfo{number}{7–8} (\bibinfo{date}{Oct.} \bibinfo{year}{1998}), \bibinfo{pages}{573–617}.
\newblock
\showISSN{0883-9514, 1087-6545}
\href{https://doi.org/10.1080/088395198117550}{doi:\nolinkurl{10.1080/088395198117550}}


\bibitem[Dey et~al\mbox{.}(2018)]%
        {dey_effects_2018}
\bibfield{author}{\bibinfo{person}{Arindam Dey}, \bibinfo{person}{Hao Chen}, \bibinfo{person}{Chang Zhuang}, \bibinfo{person}{Mark Billinghurst}, {and} \bibinfo{person}{Robert~W. Lindeman}.} \bibinfo{year}{2018}\natexlab{}.
\newblock \showarticletitle{Effects of {Sharing} {Real}-{Time} {Multi}-{Sensory} {Heart} {Rate} {Feedback} in {Different} {Immersive} {Collaborative} {Virtual} {Environments}}. In \bibinfo{booktitle}{\emph{2018 {IEEE} {International} {Symposium} on {Mixed} and {Augmented} {Reality} ({ISMAR})}}. \bibinfo{pages}{165--173}.
\newblock
\href{https://doi.org/10.1109/ISMAR.2018.00052}{doi:\nolinkurl{10.1109/ISMAR.2018.00052}}
\newblock
\shownote{ISSN: 1554-7868}.


\bibitem[Ekman(1971)]%
        {ekman_universals_1971}
\bibfield{author}{\bibinfo{person}{Paul Ekman}.} \bibinfo{year}{1971}\natexlab{}.
\newblock \showarticletitle{Universals and cultural differences in facial expressions of emotion.}. In \bibinfo{booktitle}{\emph{Nebraska symposium on motivation}}. \bibinfo{publisher}{University of Nebraska Press}.
\newblock
\urldef\tempurl%
\url{https://psycnet.apa.org/record/1973-11154-001}
\showURL{%
\tempurl}


\bibitem[Evans and Stanovich(2013)]%
        {evans2013dual}
\bibfield{author}{\bibinfo{person}{Jonathan St~BT Evans} {and} \bibinfo{person}{Keith~E Stanovich}.} \bibinfo{year}{2013}\natexlab{}.
\newblock \showarticletitle{Dual-process theories of higher cognition: Advancing the debate}.
\newblock \bibinfo{journal}{\emph{Perspectives on psychological science}} \bibinfo{volume}{8}, \bibinfo{number}{3} (\bibinfo{year}{2013}), \bibinfo{pages}{223--241}.
\newblock


\bibitem[Gong et~al\mbox{.}(2025)]%
        {gong_comparing_2025}
\bibfield{author}{\bibinfo{person}{Daojun Gong}, \bibinfo{person}{Haoming Yan}, \bibinfo{person}{Ming Wu}, \bibinfo{person}{Yimin Wang}, \bibinfo{person}{Yifu Lei}, \bibinfo{person}{Xuewen Wang}, {and} \bibinfo{person}{Ruowei Xiao}.} \bibinfo{year}{2025}\natexlab{}.
\newblock \showarticletitle{Comparing {Physiological} {Synchrony} and {User} {Copresent} {Experience} in {Virtual} {Reality}: {A} {Quantitative}–{Qualitative} {Gap}}.
\newblock \bibinfo{journal}{\emph{Electronics}} \bibinfo{volume}{14}, \bibinfo{number}{6} (\bibinfo{date}{Jan.} \bibinfo{year}{2025}), \bibinfo{pages}{1129}.
\newblock
\showISSN{2079-9292}
\href{https://doi.org/10.3390/electronics14061129}{doi:\nolinkurl{10.3390/electronics14061129}}
\newblock
\shownote{Number: 6 Publisher: Multidisciplinary Digital Publishing Institute}.


\bibitem[Haddock et~al\mbox{.}(2008)]%
        {Haddock_Maio_Arnold_Huskinson_2008}
\bibfield{author}{\bibinfo{person}{Geoffrey Haddock}, \bibinfo{person}{Gregory~R. Maio}, \bibinfo{person}{Karin Arnold}, {and} \bibinfo{person}{Thomas Huskinson}.} \bibinfo{year}{2008}\natexlab{}.
\newblock \showarticletitle{Should Persuasion Be Affective or Cognitive? The Moderating Effects of Need for Affect and Need for Cognition}.
\newblock \bibinfo{journal}{\emph{Personality and Social Psychology Bulletin}} (\bibinfo{date}{March} \bibinfo{year}{2008}).
\newblock
\href{https://doi.org/10.1177/0146167208314871}{doi:\nolinkurl{10.1177/0146167208314871}}


\bibitem[Halbig and Latoschik(2021)]%
        {halbig_systematic_2021}
\bibfield{author}{\bibinfo{person}{Andreas Halbig} {and} \bibinfo{person}{Marc~Erich Latoschik}.} \bibinfo{year}{2021}\natexlab{}.
\newblock \showarticletitle{A systematic review of physiological measurements, factors, methods, and applications in virtual reality}.
\newblock \bibinfo{journal}{\emph{Frontiers in Virtual Reality}}  \bibinfo{volume}{2} (\bibinfo{year}{2021}), \bibinfo{pages}{694567}.
\newblock
\urldef\tempurl%
\url{https://www.frontiersin.org/articles/10.3389/frvir.2021.694567/full}
\showURL{%
\tempurl}
\newblock
\shownote{Publisher: Frontiers Media SA}.


\bibitem[Harrington and Dillahunt(2021)]%
        {harrington2021eliciting}
\bibfield{author}{\bibinfo{person}{Christina Harrington} {and} \bibinfo{person}{Tawanna~R Dillahunt}.} \bibinfo{year}{2021}\natexlab{}.
\newblock \showarticletitle{Eliciting tech futures among Black young adults: A case study of remote speculative co-design}. In \bibinfo{booktitle}{\emph{Proceedings of the 2021 CHI conference on human factors in computing systems}}. \bibinfo{pages}{1--15}.
\newblock


\bibitem[He et~al\mbox{.}(2020)]%
        {He_Zhou_Liu_Hu_2020}
\bibfield{author}{\bibinfo{person}{Jing-Xian He}, \bibinfo{person}{Li Zhou}, \bibinfo{person}{Zhen-Tao Liu}, {and} \bibinfo{person}{Xin-Yue Hu}.} \bibinfo{year}{2020}\natexlab{}.
\newblock \showarticletitle{Digital Empirical Research of Influencing Factors of Musical Emotion Classification Based on Pleasure-Arousal Musical Emotion Fuzzy Model}.
\newblock \bibinfo{journal}{\emph{Journal of Advanced Computational Intelligence and Intelligent Informatics}} \bibinfo{volume}{24}, \bibinfo{number}{7} (\bibinfo{date}{Dec.} \bibinfo{year}{2020}), \bibinfo{pages}{872–881}.
\newblock
\href{https://doi.org/10.20965/jaciii.2020.p0872}{doi:\nolinkurl{10.20965/jaciii.2020.p0872}}


\bibitem[Hirsch et~al\mbox{.}(2023)]%
        {hirsch2023my}
\bibfield{author}{\bibinfo{person}{Linda Hirsch}, \bibinfo{person}{Florian M{\"u}ller}, \bibinfo{person}{Francesco Chiossi}, \bibinfo{person}{Theodor Benga}, {and} \bibinfo{person}{Andreas~Martin Butz}.} \bibinfo{year}{2023}\natexlab{}.
\newblock \showarticletitle{My heart will go on: Implicitly increasing social connectedness by visualizing asynchronous players’ heartbeats in VR games}.
\newblock \bibinfo{journal}{\emph{Proceedings of the ACM on Human-Computer Interaction}} \bibinfo{volume}{7}, \bibinfo{number}{CHI PLAY} (\bibinfo{year}{2023}), \bibinfo{pages}{976--1001}.
\newblock


\bibitem[Hove and Risen(2009)]%
        {hove2009s}
\bibfield{author}{\bibinfo{person}{Michael~J Hove} {and} \bibinfo{person}{Jane~L Risen}.} \bibinfo{year}{2009}\natexlab{}.
\newblock \showarticletitle{It's all in the timing: Interpersonal synchrony increases affiliation}.
\newblock \bibinfo{journal}{\emph{Social cognition}} \bibinfo{volume}{27}, \bibinfo{number}{6} (\bibinfo{year}{2009}), \bibinfo{pages}{949--960}.
\newblock


\bibitem[Isola et~al\mbox{.}(2017)]%
        {isola2017image}
\bibfield{author}{\bibinfo{person}{Phillip Isola}, \bibinfo{person}{Jun-Yan Zhu}, \bibinfo{person}{Tinghui Zhou}, {and} \bibinfo{person}{Alexei~A Efros}.} \bibinfo{year}{2017}\natexlab{}.
\newblock \showarticletitle{Image-to-image translation with conditional adversarial networks}. In \bibinfo{booktitle}{\emph{Proceedings of the IEEE conference on computer vision and pattern recognition}}. \bibinfo{pages}{1125--1134}.
\newblock


\bibitem[Kahneman(1973)]%
        {kahneman1973attention}
\bibfield{author}{\bibinfo{person}{Daniel Kahneman}.} \bibinfo{year}{1973}\natexlab{}.
\newblock \bibinfo{booktitle}{\emph{Attention and effort}}. Vol.~\bibinfo{volume}{1063}.
\newblock \bibinfo{publisher}{Citeseer}.
\newblock


\bibitem[Lake et~al\mbox{.}(2017)]%
        {lake2017building}
\bibfield{author}{\bibinfo{person}{Brenden~M Lake}, \bibinfo{person}{Tomer~D Ullman}, \bibinfo{person}{Joshua~B Tenenbaum}, {and} \bibinfo{person}{Samuel~J Gershman}.} \bibinfo{year}{2017}\natexlab{}.
\newblock \showarticletitle{Building machines that learn and think like people}.
\newblock \bibinfo{journal}{\emph{Behavioral and brain sciences}}  \bibinfo{volume}{40} (\bibinfo{year}{2017}), \bibinfo{pages}{e253}.
\newblock


\bibitem[Lakoff and Johnson(2008)]%
        {lakoff2008metaphors}
\bibfield{author}{\bibinfo{person}{George Lakoff} {and} \bibinfo{person}{Mark Johnson}.} \bibinfo{year}{2008}\natexlab{}.
\newblock \bibinfo{booktitle}{\emph{Metaphors we live by}}.
\newblock \bibinfo{publisher}{University of Chicago press}.
\newblock


\bibitem[Lanovaz et~al\mbox{.}(2020)]%
        {lanovaz2020machine}
\bibfield{author}{\bibinfo{person}{Marc~J Lanovaz}, \bibinfo{person}{Antonia~R Giannakakos}, {and} \bibinfo{person}{Oc{\'e}ane Destras}.} \bibinfo{year}{2020}\natexlab{}.
\newblock \showarticletitle{Machine learning to analyze single-case data: A proof of concept}.
\newblock \bibinfo{journal}{\emph{Perspectives on Behavior Science}} \bibinfo{volume}{43}, \bibinfo{number}{1} (\bibinfo{year}{2020}), \bibinfo{pages}{21--38}.
\newblock


\bibitem[Lee et~al\mbox{.}(2022)]%
        {lee_understanding_2022}
\bibfield{author}{\bibinfo{person}{Sueyoon Lee}, \bibinfo{person}{Abdallah El~Ali}, \bibinfo{person}{Maarten Wijntjes}, {and} \bibinfo{person}{Pablo Cesar}.} \bibinfo{year}{2022}\natexlab{}.
\newblock \showarticletitle{Understanding and {Designing} {Avatar} {Biosignal} {Visualizations} for {Social} {Virtual} {Reality} {Entertainment}}.
\newblock \bibinfo{journal}{\emph{CHI Conference on Human Factors in Computing Systems}} (\bibinfo{date}{April} \bibinfo{year}{2022}), \bibinfo{pages}{1--15}.
\newblock
\href{https://doi.org/10.1145/3491102.3517451}{doi:\nolinkurl{10.1145/3491102.3517451}}
\newblock
\shownote{Conference Name: CHI '22: CHI Conference on Human Factors in Computing Systems ISBN: 9781450391573 Place: New Orleans LA USA Publisher: ACM}.


\bibitem[Lux et~al\mbox{.}(2018)]%
        {lux2018live}
\bibfield{author}{\bibinfo{person}{Ewa Lux}, \bibinfo{person}{Marc~TP Adam}, \bibinfo{person}{Verena Dorner}, \bibinfo{person}{Sina Helming}, \bibinfo{person}{Michael~T Knierim}, {and} \bibinfo{person}{Christof Weinhardt}.} \bibinfo{year}{2018}\natexlab{}.
\newblock \showarticletitle{Live biofeedback as a user interface design element: A review of the literature}.
\newblock \bibinfo{journal}{\emph{Communications of the Association for Information Systems}} \bibinfo{volume}{43}, \bibinfo{number}{1} (\bibinfo{year}{2018}), \bibinfo{pages}{18}.
\newblock


\bibitem[Ma{\~n}as et~al\mbox{.}(2024)]%
        {manas2024improving}
\bibfield{author}{\bibinfo{person}{Oscar Ma{\~n}as}, \bibinfo{person}{Pietro Astolfi}, \bibinfo{person}{Melissa Hall}, \bibinfo{person}{Candace Ross}, \bibinfo{person}{Jack Urbanek}, \bibinfo{person}{Adina Williams}, \bibinfo{person}{Aishwarya Agrawal}, \bibinfo{person}{Adriana Romero-Soriano}, {and} \bibinfo{person}{Michal Drozdzal}.} \bibinfo{year}{2024}\natexlab{}.
\newblock \showarticletitle{Improving text-to-image consistency via automatic prompt optimization}.
\newblock \bibinfo{journal}{\emph{arXiv preprint arXiv:2403.17804}} (\bibinfo{year}{2024}).
\newblock


\bibitem[Mar{\'\i}n-Morales et~al\mbox{.}(2018)]%
        {marin2018affective}
\bibfield{author}{\bibinfo{person}{Javier Mar{\'\i}n-Morales}, \bibinfo{person}{Juan~Luis Higuera-Trujillo}, \bibinfo{person}{Alberto Greco}, \bibinfo{person}{Jaime Guixeres}, \bibinfo{person}{Carmen Llinares}, \bibinfo{person}{Enzo~Pasquale Scilingo}, \bibinfo{person}{Mariano Alca{\~n}iz}, {and} \bibinfo{person}{Gaetano Valenza}.} \bibinfo{year}{2018}\natexlab{}.
\newblock \showarticletitle{Affective computing in virtual reality: emotion recognition from brain and heartbeat dynamics using wearable sensors}.
\newblock \bibinfo{journal}{\emph{Scientific reports}} \bibinfo{volume}{8}, \bibinfo{number}{1} (\bibinfo{year}{2018}), \bibinfo{pages}{13657}.
\newblock


\bibitem[Marín-Morales et~al\mbox{.}(2020)]%
        {marin-morales_emotion_2020}
\bibfield{author}{\bibinfo{person}{Javier Marín-Morales}, \bibinfo{person}{Carmen Llinares}, \bibinfo{person}{Jaime Guixeres}, {and} \bibinfo{person}{Mariano Alcañiz}.} \bibinfo{year}{2020}\natexlab{}.
\newblock \showarticletitle{Emotion {Recognition} in {Immersive} {Virtual} {Reality}: {From} {Statistics} to {Affective} {Computing}}.
\newblock \bibinfo{journal}{\emph{Sensors}} \bibinfo{volume}{20}, \bibinfo{number}{18} (\bibinfo{date}{Jan.} \bibinfo{year}{2020}), \bibinfo{pages}{5163}.
\newblock
\showISSN{1424-8220}
\href{https://doi.org/10.3390/s20185163}{doi:\nolinkurl{10.3390/s20185163}}
\newblock
\shownote{Number: 18 Publisher: Multidisciplinary Digital Publishing Institute}.


\bibitem[Mateos-García et~al\mbox{.}(2023)]%
        {mateos-garcia_driver_2023}
\bibfield{author}{\bibinfo{person}{Nuria Mateos-García}, \bibinfo{person}{Ana-Belén Gil-González}, \bibinfo{person}{Ana Luis-Reboredo}, {and} \bibinfo{person}{Belén Pérez-Lancho}.} \bibinfo{year}{2023}\natexlab{}.
\newblock \showarticletitle{Driver stress detection from physiological signals by virtual reality simulator}.
\newblock \bibinfo{journal}{\emph{Electronics}} \bibinfo{volume}{12}, \bibinfo{number}{10} (\bibinfo{year}{2023}), \bibinfo{pages}{2179}.
\newblock
\urldef\tempurl%
\url{https://www.mdpi.com/2079-9292/12/10/2179}
\showURL{%
\tempurl}
\newblock
\shownote{Publisher: MDPI}.


\bibitem[Mehrabian(1996)]%
        {mehrabian1996pleasure}
\bibfield{author}{\bibinfo{person}{Albert Mehrabian}.} \bibinfo{year}{1996}\natexlab{}.
\newblock \showarticletitle{Pleasure-arousal-dominance: A general framework for describing and measuring individual differences in temperament}.
\newblock \bibinfo{journal}{\emph{Current psychology}} \bibinfo{volume}{14}, \bibinfo{number}{4} (\bibinfo{year}{1996}), \bibinfo{pages}{261--292}.
\newblock


\bibitem[Miles et~al\mbox{.}(2014)]%
        {miles2014qualitative}
\bibfield{author}{\bibinfo{person}{Matthew~B Miles}, \bibinfo{person}{A~Michael Huberman}, {and} \bibinfo{person}{Johnny Saldana}.} \bibinfo{year}{2014}\natexlab{}.
\newblock \showarticletitle{Qualitative data analysis: A methods sourcebook}.
\newblock \bibinfo{journal}{\emph{(No Title)}} (\bibinfo{year}{2014}).
\newblock


\bibitem[Moreira et~al\mbox{.}(2022)]%
        {moreira2022toward}
\bibfield{author}{\bibinfo{person}{Catarina Moreira}, \bibinfo{person}{Francisco~PM Sim{\~o}es}, \bibinfo{person}{Mark~JW Lee}, \bibinfo{person}{Ezequiel~R Zorzal}, \bibinfo{person}{Robert~W Lindeman}, \bibinfo{person}{Jo{\~a}o~Madeiras Pereira}, \bibinfo{person}{Kyle Johnsen}, {and} \bibinfo{person}{Joaquim Jorge}.} \bibinfo{year}{2022}\natexlab{}.
\newblock \showarticletitle{Toward VR in VR: assessing engagement and social interaction in a virtual conference}.
\newblock \bibinfo{journal}{\emph{IEEE Access}}  \bibinfo{volume}{11} (\bibinfo{year}{2022}), \bibinfo{pages}{1906--1922}.
\newblock


\bibitem[Mou et~al\mbox{.}(2015)]%
        {mou2015group}
\bibfield{author}{\bibinfo{person}{Wenxuan Mou}, \bibinfo{person}{Oya Celiktutan}, {and} \bibinfo{person}{Hatice Gunes}.} \bibinfo{year}{2015}\natexlab{}.
\newblock \showarticletitle{Group-level arousal and valence recognition in static images: Face, body and context}. In \bibinfo{booktitle}{\emph{2015 11th IEEE International Conference and Workshops on Automatic Face and Gesture Recognition (FG)}}, Vol.~\bibinfo{volume}{5}. IEEE, \bibinfo{pages}{1--6}.
\newblock


\bibitem[Mulvale et~al\mbox{.}(2019)]%
        {mulvale2019co}
\bibfield{author}{\bibinfo{person}{Gillian Mulvale}, \bibinfo{person}{Sandra Moll}, \bibinfo{person}{Ashleigh Miatello}, \bibinfo{person}{Louise Murray-Leung}, \bibinfo{person}{Karlie Rogerson}, {and} \bibinfo{person}{Roberto~B Sassi}.} \bibinfo{year}{2019}\natexlab{}.
\newblock \showarticletitle{Co-designing services for youth with mental health issues: novel elicitation approaches}.
\newblock \bibinfo{journal}{\emph{International Journal of Qualitative Methods}}  \bibinfo{volume}{18} (\bibinfo{year}{2019}), \bibinfo{pages}{1609406918816244}.
\newblock


\bibitem[Park et~al\mbox{.}(2023)]%
        {park2023generative}
\bibfield{author}{\bibinfo{person}{Joon~Sung Park}, \bibinfo{person}{Joseph O'Brien}, \bibinfo{person}{Carrie~Jun Cai}, \bibinfo{person}{Meredith~Ringel Morris}, \bibinfo{person}{Percy Liang}, {and} \bibinfo{person}{Michael~S Bernstein}.} \bibinfo{year}{2023}\natexlab{}.
\newblock \showarticletitle{Generative agents: Interactive simulacra of human behavior}. In \bibinfo{booktitle}{\emph{Proceedings of the 36th annual acm symposium on user interface software and technology}}. \bibinfo{pages}{1--22}.
\newblock


\bibitem[Petrescu et~al\mbox{.}(2020)]%
        {petrescu_integrating_2020}
\bibfield{author}{\bibinfo{person}{Livia Petrescu}, \bibinfo{person}{Cătălin Petrescu}, \bibinfo{person}{Oana Mitruț}, \bibinfo{person}{Gabriela Moise}, \bibinfo{person}{Alin Moldoveanu}, \bibinfo{person}{Florica Moldoveanu}, {and} \bibinfo{person}{Marius Leordeanu}.} \bibinfo{year}{2020}\natexlab{}.
\newblock \showarticletitle{Integrating biosignals measurement in virtual reality environments for anxiety detection}.
\newblock \bibinfo{journal}{\emph{Sensors}} \bibinfo{volume}{20}, \bibinfo{number}{24} (\bibinfo{year}{2020}), \bibinfo{pages}{7088}.
\newblock
\urldef\tempurl%
\url{https://www.mdpi.com/1424-8220/20/24/7088}
\showURL{%
\tempurl}
\newblock
\shownote{Publisher: MDPI}.


\bibitem[Plutchik(2003)]%
        {plutchik_emotions_2003}
\bibfield{author}{\bibinfo{person}{Robert Plutchik}.} \bibinfo{year}{2003}\natexlab{}.
\newblock \bibinfo{booktitle}{\emph{Emotions and life: {Perspectives} from psychology, biology, and evolution}}.
\newblock \bibinfo{publisher}{American Psychological Association}, \bibinfo{address}{Washington, DC, US}.
\newblock
\showISBNx{978-1-55798-949-9}
\newblock
\shownote{Pages: xix, 381}.


\bibitem[Qu et~al\mbox{.}(2024)]%
        {qu2024integration}
\bibfield{author}{\bibinfo{person}{Youzhi Qu}, \bibinfo{person}{Chen Wei}, \bibinfo{person}{Penghui Du}, \bibinfo{person}{Wenxin Che}, \bibinfo{person}{Chi Zhang}, \bibinfo{person}{Wanli Ouyang}, \bibinfo{person}{Yatao Bian}, \bibinfo{person}{Feiyang Xu}, \bibinfo{person}{Bin Hu}, \bibinfo{person}{Kai Du}, {et~al\mbox{.}}} \bibinfo{year}{2024}\natexlab{}.
\newblock \showarticletitle{Integration of cognitive tasks into artificial general intelligence test for large models}.
\newblock \bibinfo{journal}{\emph{Iscience}} \bibinfo{volume}{27}, \bibinfo{number}{4} (\bibinfo{year}{2024}).
\newblock


\bibitem[Russell(1980)]%
        {russell_circumplex_1980}
\bibfield{author}{\bibinfo{person}{James~A. Russell}.} \bibinfo{year}{1980}\natexlab{}.
\newblock \showarticletitle{A circumplex model of affect.}
\newblock \bibinfo{journal}{\emph{Journal of Personality and Social Psychology}} \bibinfo{volume}{39}, \bibinfo{number}{6} (\bibinfo{date}{Dec.} \bibinfo{year}{1980}), \bibinfo{pages}{1161--1178}.
\newblock
\showISSN{1939-1315, 0022-3514}
\href{https://doi.org/10.1037/h0077714}{doi:\nolinkurl{10.1037/h0077714}}


\bibitem[Salminen et~al\mbox{.}(2019)]%
        {salminen_evoking_2019}
\bibfield{author}{\bibinfo{person}{Mikko Salminen}, \bibinfo{person}{Simo Järvelä}, \bibinfo{person}{Antti Ruonala}, \bibinfo{person}{Ville~J. Harjunen}, \bibinfo{person}{Juho Hamari}, \bibinfo{person}{Giulio Jacucci}, {and} \bibinfo{person}{Niklas Ravaja}.} \bibinfo{year}{2019}\natexlab{}.
\newblock \showarticletitle{Evoking physiological synchrony and empathy using social {VR} with biofeedback}.
\newblock \bibinfo{journal}{\emph{IEEE Transactions on Affective Computing}} \bibinfo{volume}{13}, \bibinfo{number}{2} (\bibinfo{year}{2019}), \bibinfo{pages}{746--755}.
\newblock
\urldef\tempurl%
\url{https://ieeexplore.ieee.org/abstract/document/8930025/}
\showURL{%
\tempurl}
\newblock
\shownote{Publisher: IEEE}.


\bibitem[Shin et~al\mbox{.}(2019)]%
        {shin_uncanny_2019}
\bibfield{author}{\bibinfo{person}{Mincheol Shin}, \bibinfo{person}{Se~Jung Kim}, {and} \bibinfo{person}{Frank Biocca}.} \bibinfo{year}{2019}\natexlab{}.
\newblock \showarticletitle{The uncanny valley: {No} need for any further judgments when an avatar looks eerie}.
\newblock \bibinfo{journal}{\emph{Computers in Human Behavior}}  \bibinfo{volume}{94} (\bibinfo{date}{May} \bibinfo{year}{2019}), \bibinfo{pages}{100--109}.
\newblock
\showISSN{0747-5632}
\href{https://doi.org/10.1016/j.chb.2019.01.016}{doi:\nolinkurl{10.1016/j.chb.2019.01.016}}


\bibitem[Souza and Naves(2021)]%
        {souza2021attention}
\bibfield{author}{\bibinfo{person}{Rha{\'\i}ra Helena Caetano~e Souza} {and} \bibinfo{person}{Eduardo L{\'a}zaro~Martins Naves}.} \bibinfo{year}{2021}\natexlab{}.
\newblock \showarticletitle{Attention detection in virtual environments using EEG signals: a scoping review}.
\newblock \bibinfo{journal}{\emph{frontiers in physiology}}  \bibinfo{volume}{12} (\bibinfo{year}{2021}), \bibinfo{pages}{727840}.
\newblock


\bibitem[Sumers et~al\mbox{.}(2023)]%
        {sumers2023cognitive}
\bibfield{author}{\bibinfo{person}{Theodore Sumers}, \bibinfo{person}{Shunyu Yao}, \bibinfo{person}{Karthik Narasimhan}, {and} \bibinfo{person}{Thomas Griffiths}.} \bibinfo{year}{2023}\natexlab{}.
\newblock \showarticletitle{Cognitive architectures for language agents}.
\newblock \bibinfo{journal}{\emph{Transactions on Machine Learning Research}} (\bibinfo{year}{2023}).
\newblock


\bibitem[Valdesolo and DeSteno(2011)]%
        {valdesolo2011synchrony}
\bibfield{author}{\bibinfo{person}{Piercarlo Valdesolo} {and} \bibinfo{person}{David DeSteno}.} \bibinfo{year}{2011}\natexlab{}.
\newblock \showarticletitle{Synchrony and the social tuning of compassion.}
\newblock \bibinfo{journal}{\emph{Emotion}} \bibinfo{volume}{11}, \bibinfo{number}{2} (\bibinfo{year}{2011}), \bibinfo{pages}{262}.
\newblock


\bibitem[Wang et~al\mbox{.}(2024)]%
        {wang_systematic_2024}
\bibfield{author}{\bibinfo{person}{Yimin Wang}, \bibinfo{person}{Daojun Gong}, \bibinfo{person}{Ruowei Xiao}, \bibinfo{person}{Xinyi Wu}, {and} \bibinfo{person}{Hengbin Zhang}.} \bibinfo{year}{2024}\natexlab{}.
\newblock \showarticletitle{A {Systematic} {Review} on {Extended} {Reality}-{Mediated} {Multi}-{User} {Social} {Engagement}}.
\newblock \bibinfo{journal}{\emph{Systems}} \bibinfo{volume}{12}, \bibinfo{number}{10} (\bibinfo{date}{Sept.} \bibinfo{year}{2024}), \bibinfo{pages}{396}.
\newblock
\showISSN{2079-8954}
\href{https://doi.org/10.3390/systems12100396}{doi:\nolinkurl{10.3390/systems12100396}}


\bibitem[Wei et~al\mbox{.}(2022)]%
        {wei2022chain}
\bibfield{author}{\bibinfo{person}{Jason Wei}, \bibinfo{person}{Xuezhi Wang}, \bibinfo{person}{Dale Schuurmans}, \bibinfo{person}{Maarten Bosma}, \bibinfo{person}{Fei Xia}, \bibinfo{person}{Ed Chi}, \bibinfo{person}{Quoc~V Le}, \bibinfo{person}{Denny Zhou}, {et~al\mbox{.}}} \bibinfo{year}{2022}\natexlab{}.
\newblock \showarticletitle{Chain-of-thought prompting elicits reasoning in large language models}.
\newblock \bibinfo{journal}{\emph{Advances in neural information processing systems}}  \bibinfo{volume}{35} (\bibinfo{year}{2022}), \bibinfo{pages}{24824--24837}.
\newblock


\bibitem[Williams and Moser(2019)]%
        {williams2019art}
\bibfield{author}{\bibinfo{person}{Michael Williams} {and} \bibinfo{person}{Tami Moser}.} \bibinfo{year}{2019}\natexlab{}.
\newblock \showarticletitle{The art of coding and thematic exploration in qualitative research}.
\newblock \bibinfo{journal}{\emph{International management review}} \bibinfo{volume}{15}, \bibinfo{number}{1} (\bibinfo{year}{2019}), \bibinfo{pages}{45--55}.
\newblock


\bibitem[Winters et~al\mbox{.}(2021)]%
        {winters_can_2021}
\bibfield{author}{\bibinfo{person}{R.~Michael Winters}, \bibinfo{person}{Bruce~N. Walker}, {and} \bibinfo{person}{Grace Leslie}.} \bibinfo{year}{2021}\natexlab{}.
\newblock \showarticletitle{Can {You} {Hear} {My} {Heartbeat}?: {Hearing} an {Expressive} {Biosignal} {Elicits} {Empathy}}.
\newblock \bibinfo{journal}{\emph{Proceedings of the 2021 CHI Conference on Human Factors in Computing Systems}} (\bibinfo{date}{May} \bibinfo{year}{2021}), \bibinfo{pages}{1--11}.
\newblock
\href{https://doi.org/10.1145/3411764.3445545}{doi:\nolinkurl{10.1145/3411764.3445545}}
\newblock
\shownote{Conference Name: CHI '21: CHI Conference on Human Factors in Computing Systems ISBN: 9781450380966 Place: Yokohama Japan Publisher: ACM}.


\bibitem[Wu et~al\mbox{.}(2022)]%
        {Wu_Xiao_Sun_Zhang_Ma_He_2022}
\bibfield{author}{\bibinfo{person}{Xingjiao Wu}, \bibinfo{person}{Luwei Xiao}, \bibinfo{person}{Yixuan Sun}, \bibinfo{person}{Junhang Zhang}, \bibinfo{person}{Tianlong Ma}, {and} \bibinfo{person}{Liang He}.} \bibinfo{year}{2022}\natexlab{}.
\newblock \showarticletitle{A survey of human-in-the-loop for machine learning}.
\newblock \bibinfo{journal}{\emph{Future Generation Computer Systems}}  \bibinfo{volume}{135} (\bibinfo{date}{Oct.} \bibinfo{year}{2022}), \bibinfo{pages}{364–381}.
\newblock
\showISSN{0167-739X}
\href{https://doi.org/10.1016/j.future.2022.05.014}{doi:\nolinkurl{10.1016/j.future.2022.05.014}}


\bibitem[Yang et~al\mbox{.}(2019)]%
        {yang_distinguishing_2019}
\bibfield{author}{\bibinfo{person}{Heekyung Yang}, \bibinfo{person}{Jongdae Han}, {and} \bibinfo{person}{Kyungha Min}.} \bibinfo{year}{2019}\natexlab{}.
\newblock \showarticletitle{Distinguishing {Emotional} {Responses} to {Photographs} and {Artwork} {Using} a {Deep} {Learning}-{Based} {Approach}}.
\newblock \bibinfo{journal}{\emph{Sensors}} \bibinfo{volume}{19}, \bibinfo{number}{24} (\bibinfo{date}{Jan.} \bibinfo{year}{2019}), \bibinfo{pages}{5533}.
\newblock
\showISSN{1424-8220}
\href{https://doi.org/10.3390/s19245533}{doi:\nolinkurl{10.3390/s19245533}}
\newblock
\shownote{Number: 24 Publisher: Multidisciplinary Digital Publishing Institute}.


\end{thebibliography}

% \appendix









\end{document}